\documentclass[prl,aps,twocolumn,footinbib,superscriptaddress,notitlepage]{revtex4-2}
\usepackage[T1]{fontenc}
\usepackage{amsmath}
\usepackage{amssymb}
\usepackage{graphicx}
\usepackage{dcolumn}
\usepackage{enumerate}
\usepackage{bm}
\usepackage{xcolor}
\usepackage{comment}
\usepackage{subfigure}
\usepackage{breqn}
\usepackage{hyperref}
\hypersetup{colorlinks=true, citecolor=blue, urlcolor=blue, linkcolor=blue}

\makeatletter
\let\cat@comma@active\@empty
\makeatother

\begin{document}
\title{Exploring the dynamical interplay between mass-energy equivalence, interactions\\ and entanglement in an optical lattice clock}
\author{Anjun Chu}
\email{anjun.chu@colorado.edu}
\affiliation{JILA, NIST and Department of Physics, University of Colorado, Boulder, Colorado 80309, USA}
\affiliation{Center for Theory of Quantum Matter, University of Colorado, Boulder, Colorado 80309, USA}
\author{Victor J. Martínez-Lahuerta}
\affiliation{Institut für Theoretische Physik, Leibniz Universität Hannover, Appelstraße 2, 30167 Hannover, Germany}
\author{Maya Miklos}
\affiliation{JILA, NIST and Department of Physics, University of Colorado, Boulder, Colorado 80309, USA}
\author{Kyungtae Kim}
\affiliation{JILA, NIST and Department of Physics, University of Colorado, Boulder, Colorado 80309, USA}
\author{\\Peter Zoller}
\affiliation{Institute for Quantum Optics and Quantum Information of the Austrian Academy of Sciences, 6020 Innsbruck, Austria}
\affiliation{Institute for Theoretical Physics, University of Innsbruck, 6020 Innsbruck, Austria}
\author{Klemens Hammerer}
\affiliation{Institut für Theoretische Physik, Leibniz Universität Hannover, Appelstraße 2, 30167 Hannover, Germany}
\author{Jun Ye}
\affiliation{JILA, NIST and Department of Physics, University of Colorado, Boulder, Colorado 80309, USA}
\author{Ana Maria Rey}
\affiliation{JILA, NIST and Department of Physics, University of Colorado, Boulder, Colorado 80309, USA}
\affiliation{Center for Theory of Quantum Matter, University of Colorado, Boulder, Colorado 80309, USA}
\date{\today}

\begin{abstract}
We propose protocols that probe manifestations of the mass-energy equivalence in an optical lattice clock (OLC) interrogated with spin coherent and entangled quantum states. 
To tune and uniquely distinguish the mass-energy equivalence effects (gravitational redshift and second order Doppler shift) in such a setting, we devise a dressing protocol using an additional nuclear spin state. 
We then analyze the dynamical interplay between photon-mediated interactions and gravitational redshift and show that such interplay can lead to entanglement generation and frequency synchronization dynamics. 
In the regime where all atomic spins synchronize, we show the synchronization time depends on the initial entanglement of the state and can be used as a proxy of its metrological gain compared to a classical state.
Our work opens new possibilities for exploring the effects of general relativity on quantum coherence and entanglement in OLC experiments.
\end{abstract}

\maketitle

{\it Introduction.}---Understanding the interplay between quantum mechanics (QM) and general relativity (GR) is a fundamental quest for modern science.  
Nevertheless to date, measurements capable of genuinely witnessing this simultaneous interplay have not been realized in tabletop experiments. 
A push forward towards this milestone is becoming feasible thanks to recent improvements in precision and accuracy of atomic clocks and interferometers.
For example, matter-wave interferometers have been used for stringent tests on the equivalence principle \cite{tino2021,Fray2004,rosi2017quantum,Asenbaum2020}.
In parallel, the resolution of the gravitational redshift using spatially separated clocks has improved from meter scales \cite{Chou2010,takamoto2020} to sub millimeters \cite{Bothwell2022,zheng2023}.
Furthermore, coherence times for both clock and matter-wave interferometer are now sufficiently long \cite{Bothwell2022,Xu2019} to consider integrating clock and interferometer on a single platform.
These developments open up unique opportunities to search for new physics \cite{Safronova2018,tino2021} that could help reconcile the seemingly contradictory predictions of QM and GR.

\begin{figure}[t]
    \centering
    \includegraphics[width=8.6cm]{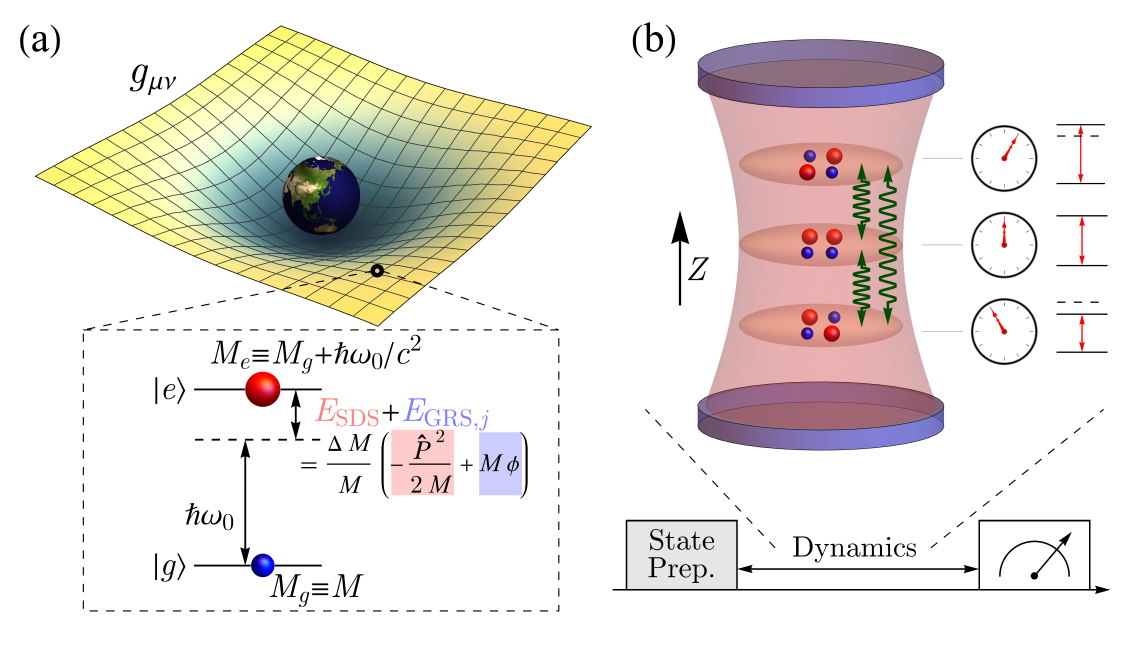}
    \caption{(a) Schematic of an optical lattice clock (OLC) embedded in the curved spacetime (metric $g_{\mu\nu}$) formed by the earth's gravity. Mass-energy equivalence is the leading order GR correction that translates internal energy difference $\hbar\omega_0$ between $|e\rangle$ and $|g\rangle$ states into a difference in the rest mass of an atom $\hbar\omega_0/c^2$. Such type of correction generates second-order Doppler shift $E_{\rm SDS}$ and gravitational redshift $E_{\mathrm{GRS},j}$ to the clock transition (see Eq.~(\ref{eq:grs}) and Eq.~(\ref{eq:sds})). (b) Schematic of probing the dynamical interplay between gravitational redshift and collective cavity-mediated interactions (see Eq.~(\ref{eq:interplay2})).}
    \label{fig:overview}
\end{figure}

Experimental developments have in parallel driven  a great  deal of theoretical effort towards the understanding of quantum dynamics with GR corrections. These progresses encompass analyses of relativistic corrections to Hamiltonians considered specifically in the context of neutral-atom and trapped-ion systems \cite{Close1970,Marzlin1995,Dimopoulos2008,Sonnleitner2018,Yudin2018,Schwartz2019,Khandelwal2020,Martinez2022,Di2021,Grochowski2021,Di2022,Werner2023}, tests of mass-energy equivalence with atoms in internal superposition states including predictions of energy-dependent phase shifts, loss of coherence and spin-motion coupling induced by gravitational time dilation \cite{Zych2011,Pikovski2015,Pikovski2017time,Zych2018,Haustein2019,Roura2020,Smith2020,Tobar2022,Asano2024}, among others \cite{Marletto2017,Bose2017,Zych2019,Paige2020,Carney2021,Pedernales2022,Christodoulou2023}. However, understanding the direct consequence of GR effects in more complex scenarios such as many-body systems, where particles can interact over the entire array, remains an outstanding problem.

In this work, we provide a first step in this direction by proposing near-term protocols to explore manifestations of single-atom GR effects, including the gravitational redshift (GRS) and the second-order Doppler shift (SDS), in the quantum many-body dynamics of an atomic ensemble. 
We take advantage of the state-of-the-art Wannier-Stark OLCs that interrogate large arrays of interacting particles under gravity \cite{Bothwell2022,Aeppli2022}.
We analyze the GRS and the SDS specifically in Wannier-Stark OLCs, both governed by the same mechanism known as mass-energy equivalence.
To distinctly characterize these GR effects and   overcome the limitation that the gravitational redshift acts just as an effective magnetic field gradient in current OLCs, we {\it firstly} devise a dressing protocol using an additional nuclear spin state to tune and uniquely distinguish the mass-energy equivalence. 
{\it Secondly}, we propose a spectroscopic protocol to probe the dynamical modification of the GRS due to photon-mediated interactions in the array.  
Depending on the relative strength of the GRS to the photon-mediated interactions, the atomic phases dynamically evolve from their individual values, dictated by the local GRS, into a semi-local or global synchronized regime. 
In both cases we observe entanglement growth from an initial product state due to the interplay of the GRS and interactions, in contrast to the entanglement generated  directly by gravitational interactions between massive objects \cite{Marletto2017,Bose2017,Carney2021,Pedernales2022,Christodoulou2023}. 
{\it Lastly}, we analyze the interplay of the GRS and interactions when they act on an initially entangled state. This is achieved  by studying the dependence of the global synchronization time on the entanglement content of the initial state. Remarkably we find the synchronization time can be used as a probe of the state's metrological utility.

\begin{figure}[t]
    \centering
    \includegraphics[width=8.6cm]{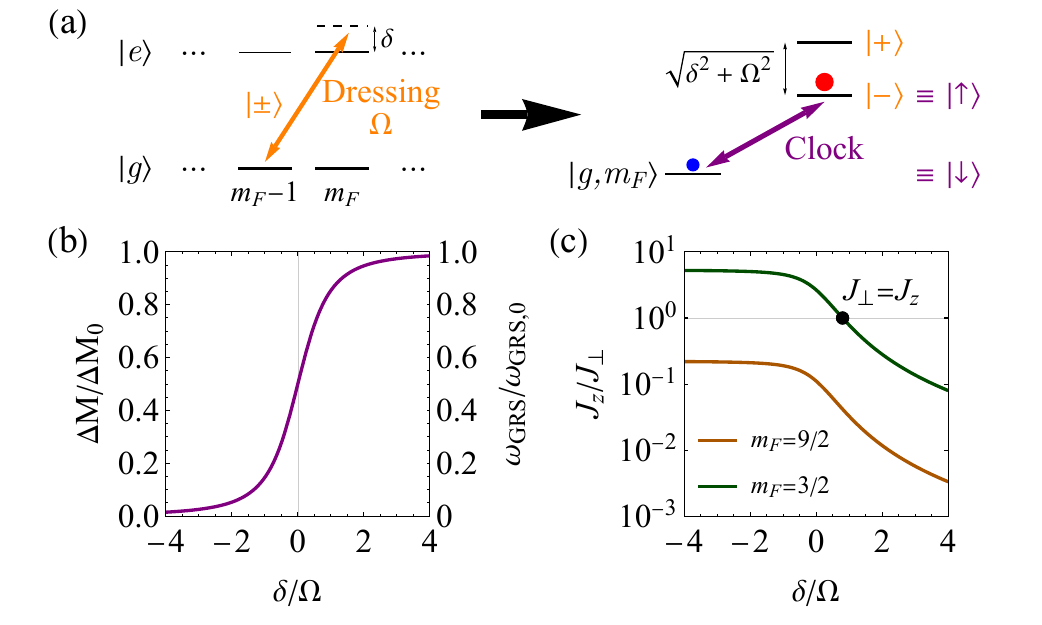}
    \caption{Tuning mass-energy equivalence via dressed states. (a) Schematic of dressing the clock transition with  another nuclear spin. The left panel show the application of the dressing laser, and the right panel show the new clock transition in the dressed basis. (b) The tunability of the mass defect $\Delta M$ and gravitational redshift $\omega_{\mathrm{GRS}}$ as a function of dressing parameter $\delta/\Omega$. $\Delta M_0$ and $\omega_{\mathrm{GRS},0}$ are the corresponding values without dressing.  (c) The tunability of cavity-mediated interactions (see $\hat{H}_{\rm cGR}$ in Eq.~(\ref{eq:interplay2})) as a function of dressing parameters $\delta/\Omega$ and nuclear spin level $m_F$. Heisenberg interaction ($J_{\perp}=J_z$) can be reached with $m_F=3/2$.}
    \label{fig:dressing}
\end{figure}

{\it Mass-energy equivalence in OLCs.}---We consider a single atom in earth's gravity described by a curved spacetime metric $g_{\mu\nu}$ (see Fig.~\ref{fig:overview}(a)).
We perform an  expansion of $g_{\mu\nu}$ in power series of $\phi/c^2$ \cite{Will2018,Note1}, with $\phi(Z)\approx g_{\rm LO} Z$ the Newtonian gravitational potential near the earth's surface and $g_{\rm LO}$ the local gravitational acceleration.
Following the treatment in Ref.~\cite{Close1970,Sonnleitner2018,Schwartz2019,Martinez2022}, one can obtain a single-atom Hamiltonian $\hat{H}_A$ accounting for the leading relativistic corrections:
\begin{equation}
    \hat{H}_A=\hat{H}_{\mathrm{point}}\bigg(M+\frac{\hat{H}_I}{c^2}\bigg) + O(c^{-4}).
    \label{eq:atom}
\end{equation}
Here, $\hat{H}_{\mathrm{point}}(M)=Mc^2+\hat{H}_0(M)+\hat{H}_{\mathrm{other}}$ is the Hamiltonian of a point particle with mass $M$,  $\hat{H}_0(M)=\hat{\mathbf{P}}^2/(2M)+M\phi$ contains the non-relativistic terms, and
$\hat{H}_{\mathrm{other}}$ contains other GR corrections in center-of-mass coordinates negligible in our case (see \footnote{See Supplemental Material at [URL will be inserted by publisher] for details of GR corrections in an OLC, additional numerical simulations of superexchange interactions, as well as analytical derivation of synchronization time, including Ref.~\cite{Will2018,Schwartz2019,Martinez2022,Werner2023,Sonnleitner2018,parker1980one,parker1982,Zhao2007,perche2021wavefunction,Alibabaei2023,norcia2018cavity}}).
The key idea of Eq.~(\ref{eq:atom}) can be understood as the mass-energy equivalence, summarized by the replacement $M\to M + \hat{H}_I/c^2$ in $\hat{H}_{\mathrm{point}}$.
OLCs feature an ultranarrow optical transition (clock transition) between two long-lived electronic states (excited state $|e\rangle$, ground state $|g\rangle$), which is described by the internal Hamiltonian $\hat{H}_I = \hbar\omega_0 |e\rangle\langle e|$, with $\omega_0$ the clock transition frequency measured at the lab position $Z=0$ (see Fig.~\ref{fig:overview}(a)).
Since in an OLC $\hat{H}_I$ contains the largest observable energy scale compared to other terms, the  mass-energy equivalence is the leading order GR correction. It  translates into a difference in the rest mass of an atom in states $|e\rangle$ and $|g\rangle$: $M_g=M$, $\Delta M_0=M_e-M_g=\hbar\omega_0/c^2$.
Note that in  general the mass defect $\Delta M=\langle \hat{H}_I\rangle/c^2$ is not simply a fixed number, and its tunability (see Fig.~\ref{fig:dressing}) is an important tool to determine the relativistic origin of the mass defect.

We assume that in OLCs atoms are trapped in a magic-wavelength 1D lattice along the gravitational potential ($Z$ axis) \cite{Bothwell2022,Aeppli2022}, where $|e\rangle$ and $|g\rangle$ states experience the same potential, $V(Z)=V_Z \sin^2(k_L Z)+M\phi$.
Here $V_{Z}$ is the lattice depth, $k_L$ is the wave number of the lattice that sets the atomic recoil energy $E_R=\hbar^2k_L^2/2M$ and lattice spacing $a_L=\pi/k_L$.
GR corrections to the optical lattice potential are negligible in our case (see \cite{Note1}).
In a tilted 1D lattice described by $V(Z)$, the motional eigenstates in the ground band are the so-called Wannier-Stark (WS) states $|W_j\rangle$, with $j$ the $Z$-lattice site index where the WS state is centered at \cite{Aeppli2022}.
Assuming the radial degrees of freedom are also  confined to the lowest ground state by an additional 2D lattice, with lattice depths $V_{X,Y}$, the eigenenergies of WS states are given by $E_j=Mg_{\rm LO}a_Lj+E_{\mathrm{band}}$, where $E_{\mathrm{band}}\approx \sum_{\alpha=X,Y,Z} E_R(\sqrt{V_{\alpha}/E_R}-1/4)$ \cite{zoller1998} is the ground band zero-point energy.
The GR corrections due to mass-energy equivalence is given by
\begin{equation}
    \hat{H}_{\mathrm{corr}}=\sum_j (E_{\mathrm{GRS},j}+E_{\mathrm{SDS}})|e,W_j\rangle\langle e,W_j|,
\end{equation}
with $E_{\mathrm{GRS},j}$ the gravitational redshift (GRS)  and $E_{\mathrm{SDS}}$ the second-order Doppler shift (SDS).
Their orders of magnitude are discussed below for $^{87}$Sr atoms.

Applying the mass-energy equivalence to the gravitational potential energy $M\phi$, we get the GRS (or gravitational time dilation) between $|e\rangle$ and $|g\rangle$ states ($\Delta M=\Delta M_0$),
\begin{equation}
    E_{\mathrm{GRS},j}=\frac{\Delta M}{M}\langle W_j|M\phi|W_j\rangle=\hbar\omega_0\frac{g_{\rm LO}a_Lj}{c^2}.
    \label{eq:grs}
\end{equation}
The GRS leads to a gradient of frequency shifts across the lattice. For example, the fractional frequency difference for nearest-neighbor lattice sites is just $4.4\times 10^{-23}$, while it is at the order of $10^{-19}$ for $1$~mm spatial separation as recently observed \cite{Bothwell2022,zheng2023}.

The contribution of mass-energy equivalence in the kinetic energy leads to a local modification of the zero-point energy known as SDS (or motional time dilation in special relativity) between $|e\rangle$ and $|g\rangle$ states,
\begin{equation}
    E_{\mathrm{SDS}}=-\frac{\Delta M}{M}\frac{\langle W_j|\hat{\mathbf{P}}^2|W_j\rangle}{2M}=-\frac{\hbar\omega_0}{2Mc^2} E_{\mathrm{band}}.
     \label{eq:sds}
\end{equation}
The magnitude of $E_{\mathrm{SDS}}$ increases with the lattice depth. 
For example, a deep lattice with $V_{X,Y,Z}=300E_R$ leads to fractional frequency shift $-4.5\times 10^{-21}$. 
Corrections in the kinetic energy can also lead to a modification of the WS wave functions for $|e\rangle$ atoms, while they play a negligible role compared to $E_{\mathrm{GRS},j}$ and $E_{\rm SDS}$.

{\it Tuning and distinguishing GR effects.}---In standard OLCs, the effects of GRS might be mimicked by a weak magnetic field gradient.
To provide further evidence of genuine GR effects beyond ruling out all possible systematics, one approach is to simultaneously observe $E_{\mathrm{GRS},j}$ and $E_{\rm SDS}$ in the same system. 
This could be achieved  in next-term OLCs by populating higher motional bands if  the  systematic uncertainty of lattice Stark shifts \cite{Kim2023} is  suppressed below $10^{-20}$. 

An alternative approach is to use dressed states as means to tune the mass defect $\Delta M$ and with it  simultaneously change $E_{\mathrm{GRS},j}$ and $E_{\rm SDS}$.
As shown in Fig.~\ref{fig:dressing}(a), we make use of the intrinsic multilevel structure in fermionic alkaline earth atoms with nuclear spin $F$.
We apply a dressing beam with Rabi frequency $\Omega$ and detuning $\delta$ connecting $|e,m_F\rangle$ with $|g,m_F-1\rangle$ states, leading to the dressed states, $|+\rangle=C_1 |e,m_F\rangle + C_2 |g,m_F-1\rangle$, $|-\rangle=-C_2 |e,m_F\rangle + C_1 |g,m_F-1\rangle$, 
with $C_1=(1-\delta/\sqrt{\Omega^2+\delta^2})^{1/2}/\sqrt{2}$, $C_2=(1+\delta/\sqrt{\Omega^2+\delta^2})^{1/2}/\sqrt{2}$ in the rotating frame of the dressing laser (see Appendix).
By addressing the transition between $|\uparrow\rangle\equiv|-\rangle$ and $|\downarrow\rangle\equiv|g,m_F\rangle$ states with a clock laser for the $\Delta m_F=0$ transition, one can therefore get a tunable mass defect $\Delta M = |C_2|^2\Delta M_0$ in the dressed clock transition via scanning the dressing parameter $\delta/\Omega$ (see Fig.~\ref{fig:dressing}(b)).  
Since the nuclear spin states in the ground manifold share the same mass $M$ but different Zeeman shifts, the dressing allows us to differentiate between a gravitational redshift and a  magnetic field gradient.  
In the lab frame, we can understand the tunability of the  mass-energy equivalence achieved by the dressing scheme by noticing that the state $|-\rangle$ has a probability $|C_2|^2$ to be in the  $|e\rangle$ level and therefore an average internal energy of $|C_2|^2\hbar\omega_0$.

This protocol is feasible thanks to the fact that the clock transitions between different nuclear spins are frequency resolved due to magnetic Zeeman shifts.
We also assume all other dynamical frequencies are smaller than the dressed state energy gap $\sqrt{\Omega^2+\delta^2}$ and the  Zeeman shifts between nuclear spins.
To guarantee the matching of laser phases for each atom, the dressing beam and the clock beam should be co-propagating. Moreover, spatial inhomogeneities in atomic detunings $\delta$ and in the dressing laser Rabi frequency $\Omega$ might obscure the effects of gravitational redshift.
For a mHz gravitational redshift arising from a cm-scale spatial separation, the spatial variations of $\delta(Z)$, $\Omega(Z)$ and other source of perturbations need to be suppressed below $10^{-4}$~Hz. 
The requirement for $\delta(Z)$ is attainable based on the parameters in Ref.~\cite{Bothwell2022}. 
The requirement for $\Omega(Z)$ could be achievable using a cavity to stabilize the spatial mode of the dressing laser.
One can also circumvent the stringent requirement for $\Omega(Z)$ by averaging the transition frequency of $|g,m_F\rangle\leftrightarrow|-\rangle$ and $|g,m_F\rangle\leftrightarrow|+\rangle$ (see Appendix), while sacrificing the tunability of mass defect ($\Delta M=\Delta M_0/2$).

{\it Many-body dynamics.---} After providing a recipe to distinguish genuine GR effects in OLCs, we further explore their manifestations in quantum many-body dynamics.
We consider photon-mediated interactions generated by placing the atoms in an optical cavity \cite{norcia2018cavity,Robinson2024}, in a regime where atomic contact interactions are controlled to be much weaker. 
The interplay between photon-mediated interactions and the GRS  is described by the following Hamiltonian (see Fig.~\ref{fig:overview}(b) and Appendix):
\begin{equation}
    \hat{H}_{\rm cGR}/\hbar=J_{\perp} \hat{\bf S}\cdot \hat{\bf S}+  (J_{z}-J_{\perp}) \hat{S}^z\hat{S}^z +\omega_{\mathrm{GRS}}\sum_j j \hat{\mathcal S}^z_j,
\label{eq:interplay2}
\end{equation} 
where $\hbar\omega_{\mathrm{GRS}}=(\Delta M)g_{\rm LO}a_L$ is the GRS between nearest neighbor sites, $J_{\perp}$ and $J_z$ are the collective exchange and Ising couplings.  
Here $\hat{\mathcal S}_j^{x,y,z}$ are collective spin operators summed over all atoms at the same height $j a_L$, and $\hat{S}^{x,y,z}=\sum_j\hat{\mathcal S}_j^{x,y,z}$. 
Based on Eq.~(\ref{eq:interplay2}), a magnetic field gradient will in principle give rise to similar single-atom inhomogeneities in the Hamiltonian, but we can tell them apart using the dressing scheme.
We drop the GR corrections for interaction terms since they are negligible in our case (see \cite{Note1}). 
While the use of a single nuclear spin state restricts the cavity exchange interactions to a single polarization mode ($J_z=0$ only), the dressing to another nuclear spin allows for coupling two polarization modes of the cavity (see Appendix), which enhances the tunability of $\hat{H}_{\rm cGR}$ and realizes collective Heisenberg interactions ($J_{\perp}=J_z$) as shown in Fig.~\ref{fig:dressing}(c).
In the following, we mainly focus on the case of $J_{\perp}=J_z$, since the $\hat{\bf S}\cdot \hat{\bf S}$ term becomes a constant and does not alter entanglement in the fully symmetric manifold.  
This requirement is unnecessary for observing frequency synchronization.

\begin{figure}[t]
    \centering
    \includegraphics[width=8.6cm]{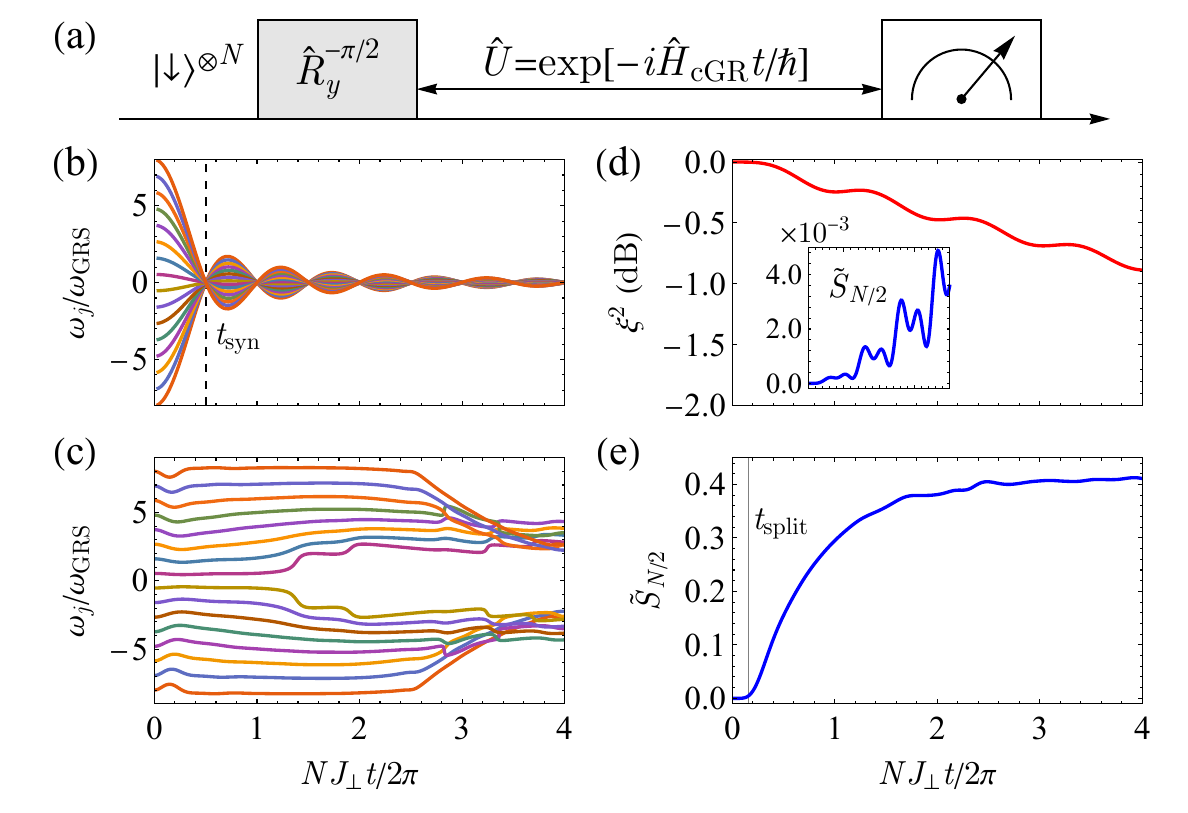}
    \caption{Interplay between photon-mediated interactions and GRS. (a) We prepare a product state with all atoms in $|\downarrow\rangle$ state and apply a laser pulse $\hat{R}_y^{-\pi/2}=\exp(i\frac{\pi}{2}\hat{S}^y)$ to start the dynamics. We focus on a single chain with $N=16$ atoms under the Hamiltonian $\hat{H}_{\rm cGR}$ (Eq.~(\ref{eq:interplay2})) with $J_{\perp}=J_z$. (b) Individual atomic frequency shift $\omega_j$ with $\omega_{\rm split}/NJ_{\perp}=0.3125$. Synchronization of atomic frequencies can be reach at time $t_{\mathrm{syn}}$. (c) Individual atomic frequency shift $\omega_j$ with $\omega_{\rm split}/NJ_{\perp}=3.125$. Global synchronization fails to occur in this regime. (d) Spin squeezing parameter $\xi^2$ and normalized Rényi entropy $\tilde{S}_{N/2}$ (inset) in the case of (b). (e) Normalized Rényi entropy $\tilde{S}_{N/2}$ in the case of (c). Entanglement starts to build up at a time scale $t_{\rm split}$ in this case.}
    \label{fig:sync}
\end{figure}

\begin{figure}[t]
    \centering
    \includegraphics[width=8.6cm]{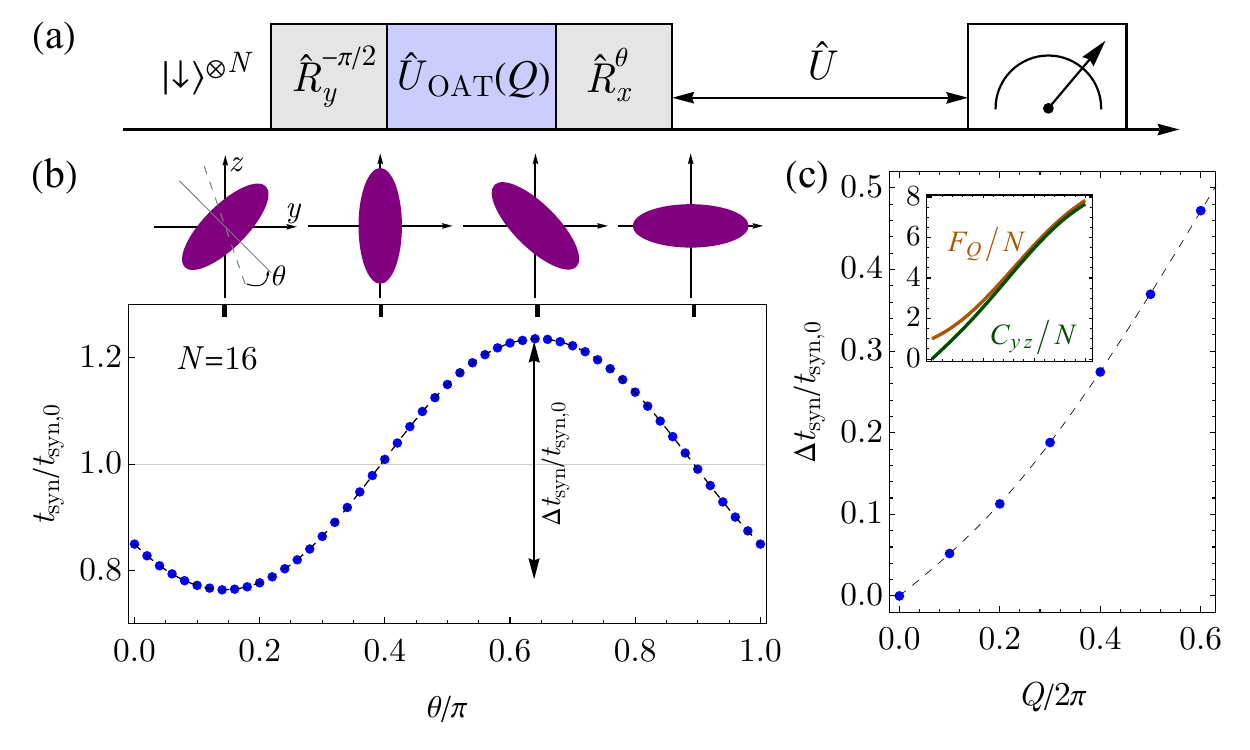}
    \caption{Interplay between  entanglement and GRS. (a) One-axis twisting (OAT) interactions $\hat{U}_{\mathrm{OAT}}(Q)=\exp(-iQ\hat{S}^z\hat{S}^z/N)$ and rotations $\hat{R}_x^{\theta}=\exp(-i\theta\hat{S}^x)$ are first applied  to generate a spin squeezed initial state (at $Q/2\pi=0.6$), followed by unitary evolution under $\hat{H}_{\rm cGR}$. (b) The synchronization time,  $t_{\mathrm{syn}}$  depends on the orientation  of the spin squeezed state  determined by  $\theta$. (c) $\Delta t_{\mathrm{syn}}$ (marked in (b)) as a function of OAT shearing strength $Q$. We show $C_{yz}=4\mathrm{max}_{\theta}\mathrm{Cov}(y,z)$ is approaching the quantum Fisher information $F_Q$ for spin squeezed states (inset). We compare the numerical simulations (blue points) for $N=16$ atoms under $\hat{H}_{\rm cGR}$ ($J_{\perp}=J_z$) with Eq.~(\ref{eq:analytic}) (black dashed lines).}
    \label{fig:squeeze}
\end{figure}

We propose to initialize all the atoms in the state $(|\uparrow\rangle+|\downarrow\rangle)/\sqrt{2}$ ($\pi/2$ pulse between the dressed clock levels), perform time evolution under Hamiltonian $\hat{H}_{\rm cGR}$ (Eq.~(\ref{eq:interplay2})), and then measure the phase shift of $\langle \hat{\mathcal{S}}^{+}_j\rangle$ for every $Z$-lattice site, resulting in frequency shift $\omega_j(t)=\tan^{-1} \left[\langle \hat{\mathcal S}^y_j\rangle/\langle \hat{\mathcal S}^x_j\rangle\right]/t$ as a function of evolution time (see Fig.~\ref{fig:sync}(a)). 
It can be observed by the application of a $\pi/2$ pulse followed by local population measurements.
Without interactions or in the case of short interrogation times, one expects to observe the GRS, $\omega_j=j\omega_{\mathrm{GRS}}$, as reported in Ref.~\cite{Bothwell2022}.
In the interaction dominated regime, the  GRS persists only for a time scale shorter than the atomic interaction time scale.
Beyond this period, the frequencies become synchronized due to interaction locking and reach $\omega_j\approx 0$ at synchronization time $t_{\mathrm{syn}}$ (see Fig.~\ref{fig:sync}(b)).
Without loss of generality, the numerical simulations in Fig.~\ref{fig:sync} and Fig.~\ref{fig:squeeze} are based on exact diagonalization for $N=16$ atoms with one atom per $Z$-lattice site.

The emergent synchronization is the result of many-body gap protection also observed in prior experiments \cite{allred_2002_prl_romalis, Deutsch:2010ky, norcia2018cavity, smale_2019_science_thywissen, davis_2020_prl_mss, young2024observing}, 
which arises when $\omega_{\rm split}\ll \Delta E$.  Here  $\omega_{\rm split}=(N_s-1)\omega_{\rm GRS}$ is the maximum redshift in the array, with $N_s$ the number of Z-lattice sites, and $\Delta E$ is the many-body gap due to Heisenberg couplings.
On the contrary, in the regime $\omega_{\rm split}\sim \Delta E$, the gap cannot maintain global synchronization (see Fig.~\ref{fig:sync}(c)). 
Using a spin wave analysis one obtains $\Delta E=NJ_{\perp}$ and $NJ_{\perp}t_{\mathrm{syn}}=\pi$ for $J_{\perp}=J_z$, where $N$ corresponds to the total atom number in the array \cite{Note1}.
For $N\sim 10^5$ $^{87}$Sr atoms, one can achieve $NJ_{\perp}/2\pi \sim$~Hz ($\omega_{\rm split}/NJ_{\perp}\sim 10^{-3}$ for cm-scale separation) \cite{norcia2018cavity}, leading to a synchronization time scale ($\sim 1$s) within reach of current experiments.

Furthermore, we find that the simultaneous presence of single-atom GRS and photon-mediated interactions can lead to quantum entanglement as shown in Fig.~\ref{fig:sync}(d,e).
In fact, in the regime $\omega_{\rm split}\ll \Delta E$, the projection of the wave function into the fully symmetric manifold imposed by the many-body gap transforms the single-particle term into an effective one-axis twisting (OAT) \cite{Kitagawa1993,ma2011} interaction term $\chi \hat{S}^z\hat{S}^z$, with $\chi\sim \omega_{\rm split}^2/[N(\Delta E)]$ (see Ref.~\cite{He2019} where the splitting is generated by a different mechanism).
In this case, entanglement builds up for $t>t_{\rm syn}$, as witnessed by a squeezing parameter \cite{ma2011}, $\xi^2\equiv \min_{\varphi}N(\Delta S_{\varphi}^{\perp})^2/|\langle\hat{\mathbf{S}}\rangle|^2 <1$ (see Fig.~\ref{fig:sync}(d)). 
Here, $(\Delta S_{\varphi}^{\perp})^2$ is the variance of spin noise along an axis perpendicular to the collective spin $\langle\hat{\mathbf{S}}\rangle$.
A faster growth of entanglement can be seen in the regime $\omega_{\rm split}\sim \Delta E$ (see Fig.~\ref{fig:sync}(e)).
Since the entanglement in this case is not captured by spin squeezing, instead we characterize the entanglement by the normalized Rényi entropy $\tilde{S}_{N/2}=-2\log_2(\mathrm{tr}(\hat{\rho}_{N/2}^2))/N$, where $\hat{\rho}_{N/2}$ is the reduced density matrix by taking partial trace over half of the system.
The entanglement builds up at a time scale $t_{\rm split}\sim \pi/\omega_{\rm split}$ in this case, which might be due to population transfer to highly entangled states in manifolds of lower total spin.
For the implementation of entanglement generation, $\omega_{\rm split}/NJ_{\perp}\sim 0.1-1$ is achievable for $10$~cm $-$ $1$~m separation and $NJ_{\perp}/2\pi\sim 0.1$~Hz. Therefore, the time scale for entanglement growth is at the order of $10$s and thus requires careful suppression of systematics.  

To study the effects of the GRS  on quantum entanglement, we consider the scenario with entangled initial states, such as the ones generated using cavity induced OAT interactions \cite{Leroux2010,pedrozo2020entanglement}, $\hat{U}_{\mathrm{OAT}}(Q)=\exp(-iQ\hat{S}^z\hat{S}^z/N)$, where $Q$ is the shearing strength (Fig.~\ref{fig:squeeze}(a)).
The squeezing direction corresponds  to the direction with minimum value of $(\Delta S_{\varphi}^{\perp})^2$, which can be controlled by performing a rotation $\hat{R}_x^{\theta}=\exp(-i\theta\hat{S}^x)$ as shown in Fig.~\ref{fig:squeeze}. 
As demonstrated in Ref.~\cite{pedrozo2020entanglement}, entangled state preparation time scales in an OLC are at the order of $10$~ms, so it is reasonable to ignore GRS in initial state preparation.
We use $|\psi_0\rangle$ to denote the state after OAT interactions, and $|\psi(\theta)\rangle$ for the state after the $\hat{R}_x^{\theta}$ rotation.

In Fig.~\ref{fig:squeeze}(b), we show that it is possible to control $t_{\mathrm{syn}}$ below or above the value of a product  initial state $t_{\mathrm{syn},0}$ ($Q=0$, as obtained in Fig.~\ref{fig:sync}) depending on the rotation $\hat{R}_x^{\theta}$.
The ratio $t_{\mathrm{syn}}/t_{\mathrm{syn},0}$ under $\hat{H}_{\rm cGR}$ ($J_{\perp}=J_z$) can be understood using the following analytic result \cite{Note1},
\begin{equation}
    \frac{t_{\mathrm{syn}}}{t_{\mathrm{syn},0}}=1-\frac{2}{\pi}\arctan\bigg[\frac{{\rm Cov}(y,z)}{(N-1)\langle\psi_0|\hat{S}^{x}|\psi_0\rangle}\bigg],
    \label{eq:analytic}
\end{equation} 
where ${\rm Cov}(\alpha,\beta)\equiv \langle\psi(\theta)|(\hat{S}^{\alpha}\hat{S}^{\beta}+\hat{S}^{\beta}\hat{S}^{\alpha})|\psi(\theta)\rangle-2\langle\psi(\theta)|\hat{S}^{\alpha}|\psi(\theta)\rangle\langle\psi(\theta)|\hat{S}^{\beta}|\psi(\theta)\rangle$, with $\alpha,\beta=x,y,z$. 
The tunability of $t_{\mathrm{syn}}$ is due to the $\theta$-dependence of ${\rm Cov}(y,z)$. 
The tunable range $\Delta t_{\mathrm{syn}}\equiv\max_{\theta} t_{\mathrm{syn}}-\min_{\theta} t_{\mathrm{syn}}$ can be used as a measure of entanglement (see Fig.~\ref{fig:squeeze}(c)), since $C_{yz}=4\max_{\theta}{\rm Cov}(y,z)$ is approaching the quantum Fisher information $F_Q$ \cite{ma2011}, which corresponds to the maximal eigenvalue of the matrix $F_{Q,\alpha\beta}=2{\rm Cov}(\alpha,\beta)$.

{\it Conclusion and outlook.}---We discussed protocols accessible in OLCs  that explore how the single-atom GR effects modify  many-body dynamics generated by photon-mediated interactions. A  similar interplay  should be observable with  atomic superexchange interactions.
While so far we have mostly focused on highly localized atomic arrays, generalizations to the case of itinerant particles where motion and other GR corrections also become relevant, will open unique opportunities for testing the basic tenets of GR when extended into the complex quantum domain.

\begin{acknowledgments}
We thank James Thompson, Raphael Kaubruegger, Dylan Young, Alexander Aeppli, Michael Werner and Torsten Zache for useful discussions.
This work is supported by the Sloan Foundation, the Simons Foundation and the Heising-Simons Foundation,  NSF JILA-PFC PHY-2317149 and NSF OMA-2016244 (QLCI) grants,  and by NIST. VML and KH acknowledge funding by the Deutsche Forschungsgemeinschaft (DFG, German Research Foundation) through Project-ID 274200144 – SFB 1227 (project A09) and Project-ID 390837967 - EXC 2123.
\end{acknowledgments}

\appendix
\section{Appendix: Derivations of the dressing scheme}
{\it Construction of dressed states.}---Here we explain the dressing scheme in Fig.~\ref{fig:dressing} in more detail. 
Due to magnetic Zeeman shifts, the clock transitions between different nuclear spins are frequency resolved, thus we can restrict the dynamics within three internal levels $|e,m_F\rangle$, $|g,m_F\rangle$ and $|g,m_F-1\rangle$. A dressing laser (with Rabi frequency $\Omega$, laser frequency $\omega_d$)  is used to couple $|e,m_F\rangle$ with $|g,m_F-1\rangle$ states (see Fig.~\ref{fig:dressing}(a)), leading to the following Hamiltonian,
\begin{equation}
    \begin{aligned}
    \hat{H}_{\rm dress}/\hbar &= \omega_0 \hat{\mathcal{P}}_{e,m_F} - \omega_Z \hat{\mathcal{P}}_{g,m_F-1}\\
    & + \frac{\Omega}{2}(|e,m_F\rangle\langle g,m_F-1|e^{-i\omega_d t}+h.c.), 
    \end{aligned}
\end{equation}
where $\omega_Z$ is the Zeeman shift between $|g,m_F\rangle$ and $|g,m_F-1\rangle$ states, and we set the energy of $|g,m_F\rangle$ state to $0$.
The projection operators are defined by $\hat{\mathcal{P}}_{\psi}=|\psi\rangle\langle \psi|$.
We then perform a unitary transformation $\hat{U}=\exp(-i\omega_d t \,\hat{\mathcal{P}}_{g,m_F-1})$ to rewrite the Hamiltonian in the rotating frame of the dressing laser,
\begin{equation}
    \begin{aligned}
    \hat{H}_{\rm dress}'/\hbar &= \omega_0 \hat{\mathcal{P}}_{e,m_F} +(\omega_0+\delta) \hat{\mathcal{P}}_{g,m_F-1}\\
    & + \frac{\Omega}{2}(|e,m_F\rangle\langle g,m_F-1|+h.c.), 
    \end{aligned}
\end{equation}
where $\delta=\omega_d-\omega_0-\omega_Z$ is the detuning of the dressing laser.
The eigenstates of this Hamiltonian are given by $|+\rangle=C_1 |e,m_F\rangle + C_2 |g,m_F-1\rangle$, $|-\rangle=-C_2 |e,m_F\rangle + C_1 |g,m_F-1\rangle$ with energy $E_{\pm}/\hbar = \omega_0+\delta/2\pm \sqrt{\Omega^2+\delta^2}/2$.
These are the dressed states defined in the main text with coefficients 
$C_1=(1-\delta/\sqrt{\Omega^2+\delta^2})^{1/2}/\sqrt{2}$ and $C_2=(1+\delta/\sqrt{\Omega^2+\delta^2})^{1/2}/\sqrt{2}$.

{\it Tunability of GR effects.}---Firstly we analyze the effects of the dressing scheme on the GR effects. We focus on the GRS described by the following Hamiltonian,
\begin{equation}
    \begin{aligned}
    \hat{H}_{\rm GRS}=(\Delta M_0)\phi\hat{\mathcal{P}}_{e,m_F}=&\,(\Delta M_0)\phi\bigg(|C_2|^2\hat{\mathcal{P}}_{-}+|C_1|^2\hat{\mathcal{P}}_{+}\\
    &+\Big(C_1C_2^{*}|+\rangle\langle -|\,+h.c.\Big)\bigg),
    \end{aligned}
\end{equation}
where $\Delta M_0=\hbar\omega_0/c^2$ is the mass defect without the dressing protocol.
As shown in Fig.~\ref{fig:dressing}(a), we probe the transition between $|g,m_F\rangle$ and $|-\rangle$. 
Consider the case that the energy gap $\sqrt{\Omega^2+\delta^2}$ between $|+\rangle$ and $|-\rangle$ states due to $\hat{H}_{\rm dress}$ is much larger than $\omega_0\phi/c^2$, the coupling between $|+\rangle$ and $|-\rangle$ states in $\hat{H}_{\rm GRS}$ are suppressed significantly. 
Therefore, to the leading order, when projected into the relevant  $|g,m_F\rangle$ and $|-\rangle$ states we can write $\hat{H}_{\mathrm{GRS}}$ as:
\begin{equation}
    \hat{H}_{\mathrm{GRS}}\approx (\Delta M_0)\phi|C_2|^2\hat{\mathcal{P}}_{-} = (\Delta M)\phi\hat{\mathcal{P}}_{-},
    \label{eq:grsm}
\end{equation}
which is equivalent to a modification of the mass defect (see Fig.~\ref{fig:dressing}(b)),
\begin{equation}
    \Delta M=|C_2|^2\Delta M_0.
    \label{eq:dressmass}
\end{equation}
The same analysis applies to the SDS.

The tunability of gravitational redshift due to the dressing scheme allows us to distinguish GRS from a magnetic field gradient. In the dressing scheme, $\hat{H}_{\rm GRS}$ leads to a position-dependent correction of the dressed state energy $E_{-}$,
\begin{equation}
    E_{-}(Z) = E_{-} + \frac{\hbar\omega_0 g_{\rm LO}}{c^2}|C_2|^2Z,
    \label{eq:dressgrs}
\end{equation} 
which can be resolved via clock spectroscopy in the effective two level system formed by $|g,m_F\rangle$ and $|-\rangle$ states.
While for a small magnetic field gradient term adding on top of a constant magnetic field, we have $\omega_0(Z) = \omega_0 + (\eta_e -\eta_g )m_F Z$, $\omega_Z(Z) = \omega_Z + \eta_g Z$, where $\eta_e=-\mathcal{G}_{^3P_0}\mu_B \partial_Z B$, $\eta_g=-\mathcal{G}_{^1S_0}\mu_B \partial_Z B$, with $\mathcal{G}_{^3P_0}$ and $\mathcal{G}_{^1S_0}$ representing the Landé g-factors, and $\mu_B$ is the Bohr magneton. So the position-dependent correction due to magnetic field gradients is given by 
\begin{equation}
    E_{-}(Z) = E_{-} + \bigg[|C_2|^2(\eta_e-\eta_g)m_F-|C_1|^2\eta_g\bigg]Z.
\end{equation}
Since $\eta_g\neq 0$, we find different dependence by varying $\delta$ compared to the GRS (see Eq.~(\ref{eq:dressgrs})). The reason is that different ground-state nuclear spins have the same mass but different Zeeman shifts.

\begin{figure}[t]
    \centering
    \includegraphics[width=8.6cm]{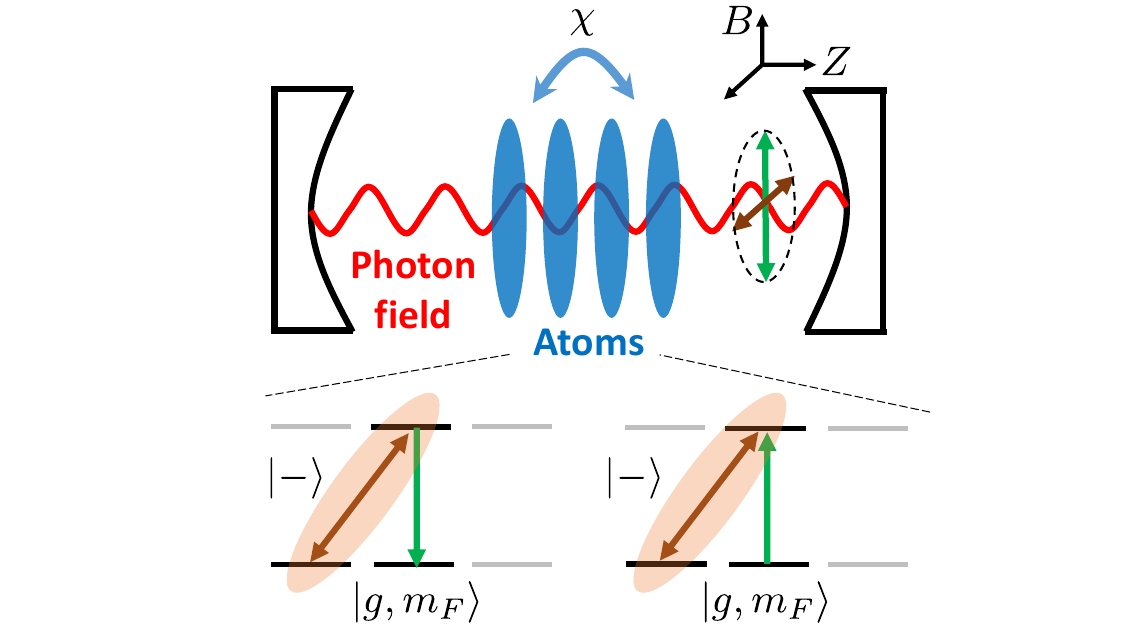}
    \caption{Schematic of the cavity-mediated interactions in the dressed basis. The cavity axis is along $Z$ and the quantization axis is along magnetic field $B$ labeled in the plot. The additional nuclear spin state in the dressing protocol allows for coupling to two polarization modes in the cavity and enhancing the tunability of the cavity-mediated interactions.}
    \label{fig:dressscheme}
\end{figure}

{\it Tunability of cavity-mediated interactions.}---Secondly we analyze the effects of the dressing protocol on the cavity-mediated interactions.
We focus on multilevel alkaline earth atoms with the quantization axis for nuclear spins perpendicular to the cavity axis (see Fig.~\ref{fig:dressscheme}).
In this case, the two polarization modes supported by the cavity can drive the $\pi$ transition and the linear combination of $\sigma^{+}$ and $\sigma^{-}$ transitions, so we can define the multilevel raising operators for these two polarization modes, $\hat{\Pi}^{+}=\sum_{jm}C^0_m |e,m\rangle_j\langle g,m|$, $\hat{\Sigma}^{+}=\sum_{jm}\frac{i}{\sqrt{2}}(C^{-1}_m |e,m-1\rangle_j\langle g,m| + C^{+1}_m |e,m+1\rangle_j\langle g,m|)$, where $j$ is the label of atoms, $m$ is the label of nuclear spins, and $C^{\sigma}_m\equiv\langle F,m; 1, \sigma|F,m+\sigma\rangle$ are the Clebsch-Gordan coefficients. 
The photon-mediated exchange interactions for multilevel alkaline earth atoms take the following form \cite{asier2022,chu2023},
\begin{equation}
    \hat{H}_c/\hbar=\chi (\hat{\Pi}^{+}\hat{\Pi}^{-}+\hat{\Sigma}^{+}\hat{\Sigma}^{-}),
\end{equation}
with $\hat{\Pi}^{-}=(\hat{\Pi}^{-})^{\dag}$ and $\hat{\Sigma}^{-}=(\hat{\Sigma}^{-})^{\dag}$.
Assuming the Zeeman shifts between nuclear spins and the energy gap $\sqrt{\Omega^2+\delta^2}$ between two dressed states are typically larger than the interaction strength $\chi N$, one can restrict the dynamics within two levels,  $|\downarrow\rangle\equiv |g,m_F\rangle$ and $|\uparrow\rangle\equiv |-\rangle$. We have
\begin{equation}
    \hat{H}_c/\hbar \approx J_{\perp} \hat{S}^{+}\hat{S}^{-} + J_z \bigg(\frac{N}{2}+\hat{S}^z\bigg)^2,
    \label{eq:cavm}
\end{equation}
where
\begin{equation}
    J_{\perp}=\chi (C^0_{m_F})^2 |C_2|^2, \quad J_z=\chi \frac{(C^{+1}_{m_F-1})^2}{2} |C_1|^2|C_2|^2.
\end{equation}
The tunability of $J_{\perp}$ and $J_z$ is shown in Fig.~\ref{fig:dressing}(c).
Eq.~(\ref{eq:interplay2}) describing the interplay between gravitational redshift and cavity-mediated interactions is a combination of Eq.~(\ref{eq:grsm}) and Eq.~(\ref{eq:cavm}) in the Wannier-Stark basis.

The physical meaning of the $J_{\perp}$ term is the spin exchange interactions between $|g,m_F\rangle$ and $|-\rangle$ state via the $\hat{\Pi}^{+}\hat{\Pi}^{-}$ process, i.e. an atom in $|-\rangle$ state emits a virtual photon into the cavity and flips to $|g,m_F\rangle$ state ($\hat{\Pi}^{-}$), and another atom in $|g,m_F\rangle$ state absorbs the same photon and flips to $|-\rangle$ state ($\hat{\Pi}^{+}$).
The physical meaning of $J_{z}$ term is the collective frequency shift of $|-\rangle$ state via the $\hat{\Sigma}^{+}\hat{\Sigma}^{-}$ process, i.e. an atom in $|-\rangle$ state absorbs a  photon from the dressing laser beam  and emits it into the cavity while staying  in the same state ($\hat{\Sigma}^{-}$), then  another atom in $|-\rangle$ state absorbs  the same cavity  photon and emits it back to the dressing laser beam ($\hat{\Sigma}^{+}$).

{\it Experimental considerations.}---Finally we focus on the experimental requirements of the dressing scheme. For $^{87}$Sr atoms, we consider the Zeeman shifts between nuclear spin states to be at the order of $10^2$~Hz. In order to frequency resolve a single transition between nuclear spin states, we have $\Omega/2\pi,\delta/2\pi\sim 10$~Hz. Considering $\chi N/2\pi$ at the order of Hz for clock transition as shown in Ref.~\cite{norcia2018cavity}, and GRS at the order of mHz for $1$~cm spatial separation, the validity of the dressing scheme is ensured.

As for experimental implementation, spatial inhomogeneities exist for the atomic frequency and the parameters $\delta$ and $\Omega$ in $E_{-}$. 
The effects of GRS will be washed out if the inhomogeneities are much larger than the value of GRS.
For $1$~cm spatial separation, the mHz scale of GRS requires to control the inhomogeneities below $10^{-4}$~Hz for direct observation.

For the atomic frequencies and detuning $\delta$, the leading order contributions are from spatial inhomogeneties in the magnetic field. One can suppress first order Zeeman shifts by probing opposite nuclear spin states and calculating the averaged frequency, which is attainable based on the parameters in Ref.~\cite{Bothwell2022}. 
This approach allows for cancellation of first order Zeeman shifts up to shot-to-shot fluctuations, and residue effects of the magnetic field can be distinguish from GRS based on the dressing scheme.

For the dressing laser Rabi frequency $\Omega$, the leading order contributions are from the spatial profile of the laser beam. If we denote the modification of $\Omega$ as $\Delta \Omega$, the change of $E_{\pm}$ is given by
\begin{equation}
    \Delta E_{\pm}/\hbar = \pm\frac{1}{2}\sqrt{\Omega^2+\delta^2}\bigg(1+\frac{(\Delta \Omega)\Omega}{\Omega^2+\delta^2}\bigg).
\end{equation}
The suppression of inhomogeneties in $E_{-}$ requires $\Delta\Omega/2\pi< 10^{-4}~$Hz, which is equivalent to $\Delta\Omega/\Omega<10^{-5}$.
In principle, this requirement is achievable using an ultrastable cavity, which allows for precise control of the spatial mode of the dressing laser.
Notice that $\Delta E_{+}+\Delta E_{-}=0$, an alternative approach to reduce this requirement is to average the transition frequency of $|g,m_F\rangle\leftrightarrow|-\rangle$ and $|g,m_F\rangle\leftrightarrow|+\rangle$,
\begin{equation}
    \frac{E_{-}(Z)+E_{+}(Z)}{2} = \hbar\omega_0+\frac{\hbar\delta}{2}+\frac{1}{2}\frac{\hbar\omega_0 }{c^2}\phi(Z).
\end{equation}
In this way, the averaged transition frequency becomes independent of $\Omega$. Even though one  sacrifices the full tunability of GRS, it is still changed to half of its  value without dressing.

\end{document}


\title{Exploring the dynamical interplay between mass-energy equivalence, interactions and entanglement in an optical lattice clock: Supplemetal Materials}
\author{Anjun Chu}
\affiliation{JILA, NIST and Department of Physics, University of Colorado, Boulder, Colorado 80309, USA}
\affiliation{Center for Theory of Quantum Matter, University of Colorado, Boulder, Colorado 80309, USA}
\author{Victor J. Martínez-Lahuerta}
\affiliation{Institut für Theoretische Physik, Leibniz Universität Hannover, Appelstraße 2, 30167 Hannover, Germany}
\author{Maya Miklos}
\affiliation{JILA, NIST and Department of Physics, University of Colorado, Boulder, Colorado 80309, USA}
\author{Kyungtae Kim}
\affiliation{JILA, NIST and Department of Physics, University of Colorado, Boulder, Colorado 80309, USA}
\author{\\Peter Zoller}
\affiliation{Institute for Quantum Optics and Quantum Information of the Austrian Academy of Sciences, 6020 Innsbruck, Austria}
\affiliation{Institute for Theoretical Physics, University of Innsbruck, 6020 Innsbruck, Austria}
\author{Klemens Hammerer}
\affiliation{Institut für Theoretische Physik, Leibniz Universität Hannover, Appelstraße 2, 30167 Hannover, Germany}
\author{Jun Ye}
\affiliation{JILA, NIST and Department of Physics, University of Colorado, Boulder, Colorado 80309, USA}
\author{Ana Maria Rey}
\affiliation{JILA, NIST and Department of Physics, University of Colorado, Boulder, Colorado 80309, USA}
\affiliation{Center for Theory of Quantum Matter, University of Colorado, Boulder, Colorado 80309, USA}
\date{\today}
\maketitle

\section{General relativistic corrections in optical lattice clocks}
Our discussion is based on the post-Newtonian expansion of Schwarzschild metric \cite{Will2018,Schwartz2019,Martinez2022,Werner2023} in the isotropic coordinates $(ct,X,Y,Z)$, 
\begin{equation}
    ds^2=g_{\mu\nu}dx^{\mu}dx^{\nu}=-\bigg(1+2\frac{\phi}{c^2}+2\frac{\phi^2}{c^4}\bigg)c^2dt^2+\bigg(1-2\frac{\phi}{c^2}\bigg)(dX^2+dY^2+dZ^2)+O\bigg(\frac{1}{c^4}\bigg),
    \label{eq:ppn}
\end{equation}
where $\phi\approx g_{\rm LO} Z$ is the Newtonian gravitational potential, with $g_{\rm LO}$ the local  gravitational acceleration. This metric defines the lab frame as it reduces to Minkowski metric at the lab position $X=Y=Z=0$.
Here we list all the relevant GR correction terms for the single-atom Hamiltonian $\hat{H}_{\rm sp}$ in the lab frame, 
\begin{equation}
    \hat{H}_{\rm sp}=\hat{H}_A+\hat{H}_{AL},
\end{equation}
and one can refer to Refs.~\cite{Schwartz2019,Martinez2022} for the details of the derivation.
$\hat{H}_A$ is the atom Hamiltonian including center of mass and internal degrees of freedom,
\begin{equation}
    \hat{H}_A=\hat{H}_{\mathrm{point}}\bigg(\hat{\mathbf{X}},\hat{\mathbf{P}},M+\frac{\hat{H}_I}{c^2}\bigg) + O\bigg(\frac{1}{c^4}\bigg),
    \label{eq:ha}
\end{equation}
where $\hat{H}_I$ is the internal Hamiltonian, and $M$ is the rest mass of a ground-state atom in center-of-mass coordinates. Here $\hat{H}_{\mathrm{point}}$ is the Hamiltonian of a point particle,
\begin{equation}
    \hat{H}_{\mathrm{point}}(\hat{\mathbf{X}},\hat{\mathbf{P}}, M)=Mc^2+\frac{\hat{\mathbf{P}}^2}{2M}+M\phi+\hat{H}_{\rm other},
    \label{eq:point}
\end{equation}
where $\hat{H}_{\rm other}$ contains the relativistic corrections,
\begin{equation}
    \hat{H}_{\rm other}=-\frac{\hat{\mathbf{P}}^4}{8M^3c^2}+M\frac{\phi^2}{2c^2}+\frac{3}{2}\frac{\hat{\mathbf{P}}\cdot\phi\hat{\mathbf{P}}}{Mc^2}+O\bigg(\frac{1}{c^4}\bigg).
    \label{eq:other}
\end{equation}
$\hat{H}_{AL}$ describes the couplings to external electromagnetic field. Although the higher order terms in multipolar expansion can play a more important role compared to the GR corrections, here we only focus on the GR corrections to the leading order electric dipole terms. Under electric dipole approximation, we have \cite{Sonnleitner2018,Schwartz2019}
\begin{equation}
    \hat{H}_{AL}=-\hat{\mathbf{d}}\cdot \mathbf{E}+\frac{1}{2M}\bigg[\hat{\mathbf{P}}\cdot(\hat{\mathbf{d}}\times \mathbf{B})+(\hat{\mathbf{d}}\times \mathbf{B})\cdot \hat{\mathbf{P}}\bigg],
    \label{eq:atomlight}
\end{equation}
where $\hat{\mathbf{d}}$ is the electric dipole moment. In $\hat{H}_{AL}$, the first term is the standard electric dipole coupling, and the second term is a relativistic correction term known as the R$\mathrm{\ddot{o}}$ntgen term, describing the coupling between a moving dipole and a magnetic field. 
The electromagnetic field in $\hat{H}_{AL}$ also contains GR corrections due to the metric in the lab frame, as pointed out in Ref.~\cite{Werner2023}. 
For example, the electric field of a plane wave propagating along $\vec{Z}$ direction (electric field polarized along $\vec{X}$ direction) should be corrected in the following way,
    \begin{equation}
        \mathbf{E}\propto\bigg(\bigg(1-2\frac{\phi}{c^2}\bigg)\exp\bigg[-i\omega t \pm \bigg(1-\frac{\phi}{c^2}\bigg)ikZ\bigg]+\mathrm{c.c.},0,0\bigg),
        \label{eq:elec}
    \end{equation}
where $k=\omega/c$. In the following, we would like to analyze the order of magnitude for all these GR corrections in optical lattice clocks.

\subsection{Atomic center-of-mass coordinates}
GR corrections in the atomic center-of-mass coordinates are given by
\begin{enumerate}
    \item Mass-energy equivalence. As shown in Eq.~(\ref{eq:ha}), the atom Hamiltonian $\hat{H}_A$ can be generated by replacing mass $M$ by $M+\hat{H}_I/c^2$ in the point-particle Hamiltonian $\hat{H}_{\rm point}$. In other words, the mass of a composite particle comprises the rest masses of the constituent particles as well as the internal energy.
    The replacement in the kinetic energy term $\hat{\mathbf{P}}^2/2M$ can lead to second-order Doppler shift (also known as motional time dilation in special relativity),
    \begin{equation}
        \hat{H}_{\rm SDS} = -\frac{\hat{\mathbf{P}}^2}{2M}\frac{\hat{H}_I}{Mc^2}.
        \label{eq:smsds}
    \end{equation}
    The replacement in the gravitational potential energy term $M\phi$ can lead to gravitational redshift (also known as gravitational time dilation),
    \begin{equation}
        \hat{H}_{\rm GRS} = \frac{\phi}{c^2}\hat{H}_I.
        \label{eq:smgrs}
    \end{equation}
    We have discussed these two types of corrections in the main text, since they are the dominant GR corrections for optical lattice clocks. 
    
    \item $\hat{H}_{\rm other}$ (see Eq.~(\ref{eq:other})). Compared to mass-energy equivalence, $\hat{H}_{\rm other}$ only acts on center-of-mass degrees of freedom and does not lead to frequency shifts of the clock transition. The observation of these corrections requires coherent superposition of wave packets at different positions, which is beyond the scope of our manuscript.
    For completeness, we still analyze their orders of magnitude in an optical lattice clock.

    In a relatively deep  optical lattice, to the leading order we can  approximate each lattice sites as an harmonic oscillator with trapping frequencies $\omega_{\alpha}=2\sqrt{V_{\alpha}E_R}/\hbar$, with $\alpha=X,Y,Z$ and $E_R$ the recoil energy. In the ground band of the optical lattice, we have
\begin{equation}
    \bigg\langle\frac{\hat{\bf P}^2}{2M}\bigg\rangle \approx \sum_{\alpha}\frac{\hbar\omega_{\alpha}}{4}, \quad \bigg\langle\frac{\hat{\bf P}^4}{8M^3c^2}\bigg\rangle \approx \sum_{\alpha}\frac{(\hbar\omega_{\alpha})^2}{16Mc^2} + \sum_{\alpha\alpha'}\frac{(\hbar\omega_{\alpha})(\hbar\omega_{\alpha'})}{32Mc^2}.
\end{equation}
For $V_{X,Y,Z}=300E_R$, we have $\langle\hat{\bf P}^4/8M^3c^2\rangle \sim 7.9\times 10^{-31}$ in fractional frequency unit.
Since  the Wannier-Stark states are localized within very few lattice sites, it is reasonable to consider $Z$ as a c-number, leading to $\langle M\phi^2/2c^2\rangle \sim 2.7\times 10^{-26}$ with $1$cm separation, and $2.7\times 10^{-22}$ with $1$m separation.
Similarly, we have $\langle \hat{\mathbf{P}}\cdot\phi\hat{\mathbf{P}}/Mc^2\rangle \sim 3.0\times 10^{-28}$ with $1$cm separation, and $3.0\times 10^{-26}$ with $1$m separation. 
These are tiny corrections compared to the mass-energy equivalence.

\end{enumerate}

\subsection{Atomic internal coordinates}
GR corrections in the atomic internal coordinates ($\hat{H}_I$) are given by
\begin{enumerate}
    \item Special relativistic corrections. These are indeed the leading order corrections of $\hat{H}_I$ based on the Dirac equation, but they are already included in the atomic physics calculations that give us information of the intrinsic atomic structure of the atom in consideration. 
    For example, the ground and excited states we considered in the manuscript are the $^{1}S_0$ and $^{3}P_0$ states of a Sr atom, and these  include corrections  due to the spin-orbit coupling effect.
    Basically, they have been included already  in the determination of the optical transition frequency $\omega_0$.

    \item Gravitational corrections. These corrections can be estimated by extending the Dirac equation into curved spacetime, and one can refer to Refs.~\cite{parker1980one,parker1982,Zhao2007,perche2021wavefunction,Alibabaei2023}. 
    For a one-electron atom, the frequency shift due to gravity is typically at the order of $a_0r_s/R^2$ with respect to the electron rest mass (see Ref.~\cite{perche2021wavefunction}), where $a_0$ is the Bohr radius, $r_s$ is the Schwarzschild radius of the earth, and $R$ is the coordinate radius with respect to the gravitational center. 
    If  the same results apply to Sr atoms, then  the frequency shift will be  typically below $10^{-23}$ in fractional frequency units. 
    These are tiny corrections compared to the mass-energy equivalence.
\end{enumerate}

\subsection{Optical lattice potential}
GR corrections of the optical lattice potential are given by
\begin{enumerate}
    \item Special relativistic correction (R$\mathrm{\ddot{o}}$ntgen term). We consider monochromatic light field in the following form, $\mathbf{E}=\mathbf{E}^{(+)}e^{-i\omega t}+\mathrm{H.c.}$, $\mathbf{B}=\mathbf{B}^{(+)}e^{-i\omega t}+\mathrm{H.c.}$, where $\omega$ is the laser frequency, and assume the lattice AC Stark shift is dominated by a single dipole-allowed transition $|a\rangle\rightarrow |b\rangle$ for simplicity, so the AC Stark shift for internal state $|a\rangle$ is given by second-order perturbation theory,
    \begin{equation}
        \hat{H}_{\rm ac}\approx -\frac{\langle a|\hat{H}_{AL}|b\rangle\langle b|\hat{H}_{AL}|a\rangle}{E_b-E_a-\hbar\omega_L},
        \label{eq:latticestark}
    \end{equation}
    where $\omega_L$ is the frequency of the lattice beam, $E_a, E_b$ are the energy of states $|a\rangle$ and $|b\rangle$ respectively.
    Here we assume linear polarization of the lattice beam, with $\mathbf{E}^{(+)}=\mathcal{E}\cos(k_LZ)\vec{X}$, $\mathbf{B}^{(+)}=(i\mathcal{E}/c)\sin(k_LZ)\vec{Y}$ and propagation direction $\vec{Z}$. Here the vector symbol $\vec{X}$ means a unit vector along $X$ direction. Plugging in Eq.~(\ref{eq:latticestark}), we have
    \begin{equation}
        \hat{H}_{\rm ac} \approx -\alpha_{\rm E1}\mathcal{E}^2\cos^2(k_LZ)+\alpha_{\rm E1}\mathcal{E}^2\frac{\hbar\omega_L}{Mc^2}\sin^2(k_LZ),
    \end{equation}
    where $\alpha_{\rm E1}=\langle a|\hat{\mathbf{d}}\cdot \vec{X}|b\rangle\langle b|\hat{\mathbf{d}}\cdot \vec{X}|a\rangle/(E_b-E_a-\hbar\omega)$ is the electric dipole polarizability. Here, the first term is the standard optical lattice potential, and the second term is the corrections due to the R$\mathrm{\ddot{o}}$ntgen term. This leads to a correction of the lattice depth at the order of $\hbar\omega_L/Mc^2\sim 10^{-11}$, which is equivalent to a fractional frequency of $10^{-20}$ for a MHz trapping potential. 
    However, this term does not lead to frequency shifts of the clock transition in a magic wavelength lattice, which shares the same $\alpha_{\rm E1}$ for the ground and excited states of the clock transition. 

    \item Gravitational correction to electromagnetic waves. As shown in Eq.~(\ref{eq:elec}), we have $\phi/c^2$ corrections in the amplitude and phase of the electric field. 
    Note that this term does not lead to differential AC Stark shifts, while it leads to a spatially dependent lattice depth and lattice spacing. 
    Considering a MHz trapping potential, such type of correction is at the order of $10^{-27}$ for $1$cm separation and $10^{-25}$ for $1$m separation in fractional frequency units. 

\end{enumerate}

Therefore, we can ignore the GR corrections of the optical lattice potential to the leading order in our case.

\subsection{Photon-mediated interactions}
Photon-mediated interactions (e.g. collective spin exchange interaction $\hat{S}^{+}\hat{S}^{-}$) originate from adiabatic elimination of cavity photons in the Tavis-Cumming Hamiltonian,
\begin{equation}
    \hat{H}_{\rm TC}/\hbar = -\Delta_c \hat{a}^{\dag}\hat{a} + g_c (\hat{a}\hat{S}^{+}+\hat{a}^{\dag}\hat{S}^{-}), 
\end{equation}
where $\hat{a}$ is annihilation operator of the cavity mode, $\Delta_c=\omega_0-\omega_c$ is the detuning between atomic transition frequency $\omega_0$ and cavity resonance $\omega_c$, and $2g_c$ is the single-photon Rabi frequency describing the strength of atom-cavity coupling. For simplicity, we assume homogeneous atom-cavity coupling among all the atoms. As for GR corrections, it would be simpler to discuss them in the Tavis-Cumming Hamiltonian $\hat{H}_{\rm TC}$.

\begin{enumerate}
    \item Special relativistic correction (R$\mathrm{\ddot{o}}$ntgen term). Plug in $\hat{H}_{\rm AL}$ the quantized electromagnetic fields of a standing-wave cavity mode, $\hat{\mathbf{E}}^{(+)} = \sqrt{\hbar\omega_c/2\epsilon_0 V}\cos(k_c Z)\hat{a} \vec{X}$, $\hat{\mathbf{B}}^{(+)} = (i/c)\sqrt{\hbar\omega_c/2\epsilon_0 V}\sin(k_c Z)\hat{a} \vec{Y}$, where $V$ is the mode volume. The atom-cavity coupling is given by
    \begin{equation}
        g_c = -\sqrt{\frac{\hbar\omega_c}{2\epsilon_0 V}} \frac{\langle e|\hat{\mathbf{d}}\cdot\vec{X}| g\rangle}{\hbar}\bigg[\int\mathrm{d}^3\mathbf{R}\;\psi^{*}_e\cos(k_c Z)\psi_g -\frac{i}{2Mc}\int\mathrm{d}^3\mathbf{R}\;\psi^{*}_e\Big(\hat{P}_Z\sin(k_c Z)+\sin(k_c Z)\hat{P}_Z\Big)\psi_g\bigg].
        \label{eq:atomlight}
    \end{equation}
Therefore, GR corrections of $g_c$ are at the order of $\langle \hat{P}_Z\rangle/Mc \sim 10^{-11}$. If we consider the cavity mode is coupled to the carrier transition of the atoms without changing the motional degrees of freedom (motional wave function $\psi_e\approx \psi_g$),  GR corrections would be further suppressed since $\hat{P}_Z\sin(k_c Z)+\sin(k_c Z)\hat{P}_Z$ is off-diagonal under a  harmonic approximation of the optical lattice. Considering photon-mediated interaction at the order of Hz \cite{norcia2018cavity}, GR corrections due to R$\mathrm{\ddot{o}}$ntgen term would be $10^{-26}$ or smaller in fractional frequency units.

    \item Gravitational correction to cavity mode function. As shown in Eq.~(\ref{eq:elec}), we have $\phi/c^2$ corrections in the cavity mode function, leading to corrections of the overlap integral in Eq.~(\ref{eq:atomlight}) as well as the cavity resonance $\omega_c$ due to the change of round-trip phase. However, similar to the discussion of the optical lattice potential, the $\phi/c^2$ corrections in cavity-mediated interactions are negligible due to the small spatial separation of the optical clocks. For $1$m spatial separation, we have $\phi/c^2\sim 10^{-16}$, leading to corrections in the  atom-cavity coupling below $10^{-31}$ in fractional frequency units.

\end{enumerate}

Therefore, we can ignore the GR corrections of the photon-mediated interactions to the leading order in our case.

\section{Detailed analysis of emergent  synchronization}
In the main text, we discuss the interplay between atomic interactions, entanglement and gravitational redshift in Fig.~3 and Fig.~4. Here we would like to provide analytic spin-wave calculations based on two different approaches, including Holstein–Primakoff approximation and restriction of dynamics within collective and spin-wave manifold, and finally discuss the experimental considerations of our protocol.

First we would like to provide numerical results for the validity of the spin-wave calculation. As shown in Fig.~\ref{fig:syntime}(a,b), we calculate the expectation value $\langle \hat{\mathbf{S}}\cdot \hat{\mathbf{S}}\rangle$ as a function of the evolution time, using the same time sequence as Fig.~3 in the main text.
Note that in a given total spin-$S$ manifold, we have $\langle \hat{\mathbf{S}}\cdot \hat{\mathbf{S}}\rangle=S(S+1)$. So $\langle \hat{\mathbf{S}}\cdot \hat{\mathbf{S}}\rangle$ can serve as a measure of population in the collective manifold ($S=N/2$), spin-wave manifold ($S=N/2-1$), as well as the manifolds with lower total spin. We consider cavity-mediated interaction $\hat{H}_{\rm cav}$ with $J_{\perp}=J_z$ (see Eq.~(5) in the main text). 
In Fig.~\ref{fig:syntime}(a), we set $\omega_{\rm split}/NJ_{\perp}=0.3125$, which is the same parameter for global frequency synchronization shown in Fig.~3(b,d) in the main text. In this case we find undamped oscillations between the collective manifold ($S=N/2$) and the spin-wave manifold ($S=N/2-1$), ensuring the validity of the spin-wave calculation.
In Fig.~\ref{fig:syntime}(b), we set $\omega_{\rm split}/NJ_{\perp}=3.125$, which is the same parameter for Fig.~3(c,e) in the main text, where global frequency synchronization fails to occur.
In this case the system can evolve to manifolds with lower total spin, so the spin-wave calculation no longer valid.
The fast growth of entanglement shown in Fig.~3(e) in the main text might due to the population in the highly entangled states in manifolds with lower total spin.

In the following, we focus on the regime of global frequency synchronization. We define $\hat{\mathcal S}_j^{x,y,z}$ as collective spin operators summed over all atoms at the same height $j a_L$, and consider $N_s$ lattice sites in $Z$ direction. 
Based on our protocol in Fig.~3 and Fig.~4 in the main text, the frequency of each atom $\omega_j(t)$ can be estimated by
\begin{equation}
    \omega_j(t) = \frac{1}{t}\arctan\bigg(\frac{\langle \hat{\mathcal{S}}^y_j(t)\rangle}{\langle \hat{\mathcal{S}}^x_j(t)\rangle}\bigg).
\end{equation}
The rigorous definition of the synchronization time $t_{\mathrm{syn}}$ is the time for the first minimum in the variance of atomic frequencies.
In fact, we can approximately reach the zero crossing of $\omega_j(t)$ for all the $j$ at $t_{\mathrm{syn}}$, as demonstrated by the analytic results below and by numerical results shown in Fig.~3(b) in the main text.

\subsection{Holstein–Primakoff approximation}
We would like to use the Holstein–Primakoff approximation to describe the case with an unentangled initial state (see Fig.~3 in the main text).
We consider $N=N_{\perp}N_s$ spins with all the spins pointing to $+x$ direction initially, where $\mathcal{S}=N_{\perp}/2$ is the total spin at the same height. 
Based on the Holstein–Primakoff approximation, we have
\begin{equation}
    \hat{\mathcal{S}}_j^x = \frac{N_{\perp}}{2}-\hat{a}_j^{\dag}\hat{a}_j, \quad \hat{\mathcal{S}}_j^y \approx \sqrt{N_{\perp}}\frac{\hat{a}_j+\hat{a}_j^{\dag}}{2}, \quad \hat{\mathcal{S}}_j^z \approx \sqrt{N_{\perp}}\frac{\hat{a}_j-\hat{a}_j^{\dag}}{2i}.
\end{equation}
In this way, the initial state becomes the vacuum state of all these bosonic operators. In the following, we will plug these bosonic operators into the Hamiltonian and keep the terms up to quadratic order of bosonic operators. We then apply Fourier transform to obtain the bosonic operators for spin waves $k=2\pi m/N_s$ with $m=0,1,2,\cdots,N_s-1$,
\begin{equation}
    \hat{a}_j=\frac{1}{\sqrt{N_s}}\sum_k e^{ikj}\hat{a}_k, \quad \hat{a}_j^{\dag}=\frac{1}{\sqrt{N_s}}\sum_k e^{-ikj}\hat{a}_k^{\dag},
\end{equation}
and rewrite the Hamiltonian accordingly. The validity of Holstein–Primakoff approximation requires $\langle \hat{a}_j\rangle \ll 1$ for all $j$. If we define $\eta=\omega_{\rm split}/(NJ_{\perp})$ as the ratio between the maximum redshift in the array ($\omega_{\rm split}=(N_s-1)\omega_{\mathrm{GRS}}$) and the spin wave excitation gap ($NJ_{\perp}$ as discussed below), the typical condition for validity would be $\eta \ll 1$.
In this regime, we have $\langle \hat{\mathcal{S}}^x_j\rangle\approx N_{\perp}/2$, so the frequency of each atom $\omega_j(t)$ can be approximated as
\begin{equation}
    \omega_j(t) \approx \frac{2\langle \hat{\mathcal{S}}^y_j(t)\rangle}{N_{\perp}t}.
\end{equation}

Here we consider the following Hamiltonian describing the interplay between collective Heisenberg interactions and gravitational redshift (see Eq.~(5) in the main text),
\begin{equation}
    \hat{H}_{\rm cGR}/\hbar=J_{\perp}\sum_{i=0}^{N_s-1}\sum_{j=0}^{N_s-1}\hat{\bf \mathcal{S}}_i\cdot\hat{\bf \mathcal{S}}_{j}+\omega_{\mathrm{GRS}}\sum_{j=0}^{N_s-1} \bigg(j-\frac{N_s-1}{2}\bigg)\hat{\bf \mathcal{S}}^z_j.
\end{equation}
Applying the Holstein–Primakoff bosons and keeping the terms up to quadratic order, the Hamiltonian in terms of spin-wave operators  becomes
\begin{equation}
    \hat{H}_{\rm cGR}/\hbar \approx -NJ_{\perp}\sum_{k \neq 0} \hat{a}_k^{\dag}\hat{a}_k + \frac{\omega_{\mathrm{GRS}}}{2i}\sqrt{N}\sum_{k\neq 0}\bigg(\frac{1}{e^{ik}-1}\hat{a}_k-\frac{1}{e^{-ik}-1}\hat{a}_k^{\dag}\bigg).
\end{equation}
Solving the Heisenberg equation of motion for $\hat{a}_k$ with $k\neq 0$, one can finally reach
\begin{equation}
    \omega_j(t)\approx -\frac{\omega_{\mathrm{GRS}}}{2}\frac{\sin\Big(NJ_{\perp}t\Big)}{NJ_{\perp}t}\sum_{k\neq 0}\frac{\sin\Big(kj+k/2\Big)}{\sin(k/2)}=\omega_{\mathrm{GRS}}\frac{\sin\Big(NJ_{\perp}t\Big)}{NJ_{\perp}t}\bigg(j-\frac{N_s-1}{2}\bigg).
    \label{eq:spw2}
\end{equation}

\begin{figure}[t]
    \centering
    \includegraphics[width=17.6cm]{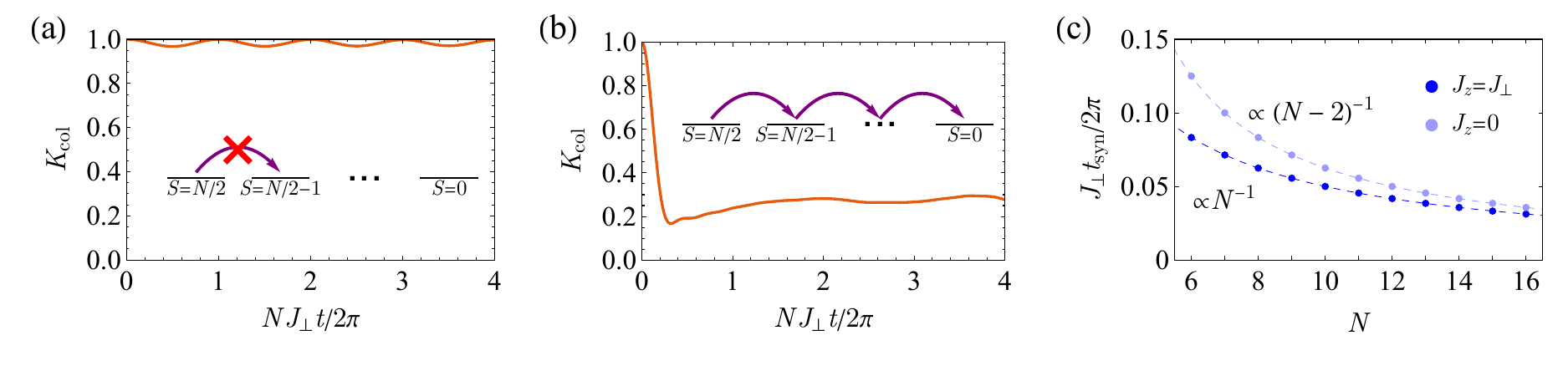}
    \caption{(a) We evolve the system based on the time sequence in Fig.~3 in the main text and calculate the expectation value $K_{\rm col}=\langle \hat{\mathbf{S}}\cdot \hat{\mathbf{S}}\rangle/[\frac{N}{2}(\frac{N}{2}+1)]$ as a function of evolution time.
    We consider Hamiltonian $\hat{H}_{\rm cGR}$ with $J_{\perp}=J_z$ (Eq.~(5) in the main text). The choice of parameters is the same as Fig.~3(b,d) in the main text, $\omega_{\rm split}/NJ_{\perp}=0.3125$. (b) The same calculation as (a) with $\omega_{\rm split}/NJ_{\perp}=3.125$, which is the same as Fig.~3(c,e) in the main text. (c) Scaling of synchronization time $t_{\rm syn}$ as a function of atom number $N$ for cavity-mediated interactions. We compare the $t_{\mathrm{syn}}$ between collective Heisenberg interactions ($J_z=J_{\perp}$) and spin exchange interactions ($J_z=0$). All the calculations are based on an initial unentangled state with all the spins pointing to $+x$ direction.}
    \label{fig:syntime}
\end{figure}

Similarly, $\omega_j(0)$ agrees with the gravitational redshift value without interactions. As for the frequency synchronization, it happens at $NJ_{\perp}t=\pi$, which gives
\begin{equation}
    J_{\perp}t_{\mathrm{syn}}=\frac{\pi}{N}.
\end{equation}
This analytic result agrees with the numerical simulations (see Fig.~\ref{fig:syntime}(b)). Within the Holstein–Primakoff approximation, there is no deviation of $\omega_j(t_{\mathrm{syn}})$ from $0$. 

\subsection{Restriction within collective and spin-wave manifold for two large spins}
For the case with an entangled initial state, it is not possible to use the Holstein–Primakoff approximation. Alternatively, we would like to simplify the Hamiltonian into two large spins ($S_1=S_2=N/4$) with effective Heisenberg interaction strength $J_{\mathrm{eff}}$ and effective redshift value $\omega_{\mathrm{eff}}$,
\begin{equation}
    \hat{H}_{\mathrm{eff}}/\hbar=2J_{\mathrm{eff}}\hat{\mathbf{S}}_1\cdot\hat{\mathbf{S}}_{2}+\frac{\omega_{\mathrm{eff}}}{2}(\hat{S}^z_1-\hat{S}^z_2).
\end{equation}
In the regime $\omega_{\mathrm{eff}} \ll NJ_{\mathrm{eff}}$, we can restrict the dynamics within the collective manifold with total spin $S=N/2$ as well as the spin-wave manifold with total spin $S=N/2-1$. Based on the Clebsch-Gordan coefficients, the states in the collective manifold ($S=N/2$) can be expressed as
\begin{equation}
    |N/2,m\rangle = \sum_{m_1} \sqrt{\frac{\binom{N/2}{N/4+m_1}\binom{N/2}{N/4+m-m_1}}{\binom{N}{N/2+m}}} |N/4,m_1\rangle_1|N/4,m-m_1\rangle_2,
    \label{eq:cg1}
\end{equation}
and the states in the spin-wave manifold ($S=N/2-1$) can be expressed as
\begin{equation}
    |N/2-1,m\rangle = \sum_{m_1} (2m_1-m)\sqrt{\frac{\binom{N/2}{N/4+m_1}\binom{N/2}{N/4+m-m_1}}{N\binom{N-2}{N/2-1+m}}} |N/4,m_1\rangle_1|N/4,m-m_1\rangle_2,
    \label{eq:cg2}
\end{equation}
where $\binom{n}{k}$ are binomial coefficients. A key observation from Eq.~(\ref{eq:cg1}) and Eq.~(\ref{eq:cg2}) is that
\begin{equation}
    (\hat{S}_1^z-\hat{S}_2^z)|N/2,m\rangle = \sqrt{\frac{(N/2+m)(N/2-m)}{N-1}} |N/2-1,m\rangle.
\end{equation}
If we restrict the dynamics within the collective and spin-wave manifold, $\hat{H}_{\mathrm{eff}}$ can be reduced to $2\times 2$ matrices in each $m$ sector in the basis $\{|N/2,m\rangle,|N/2-1,m\rangle\}$,
\begin{equation}
    \hat{H}_{\mathrm{eff},m}/\hbar=\begin{pmatrix}
       \displaystyle NJ_{\mathrm{eff}} && \displaystyle \frac{\omega_{\mathrm{eff}}}{2}\sqrt{\frac{(N/2+m)(N/2-m)}{N-1}}\\
       \displaystyle \frac{\omega_{\mathrm{eff}}}{2}\sqrt{\frac{(N/2+m)(N/2-m)}{N-1}} && 0\\
    \end{pmatrix}.
\end{equation}

Also based on Eq.~(\ref{eq:cg1}) and Eq.~(\ref{eq:cg2}), we have ($j=1,2$)
\begin{equation}
    \hat{S}_j^+|N/2,m\rangle = \frac{1}{2}\sqrt{(N/2+m+1)(N/2-m)}|N/2,m+1\rangle - \frac{(-1)^j}{2}\sqrt{\frac{(N/2-m)(N/2-m-1)}{N-1}}|N/2-1,m+1\rangle,
\end{equation}
\begin{equation}
    \hat{S}_j^-|N/2,m\rangle = \frac{1}{2}\sqrt{(N/2-m+1)(N/2+m)}|N/2,m-1\rangle + \frac{(-1)^j}{2}\sqrt{\frac{(N/2+m)(N/2+m-1)}{N-1}}|N/2-1,m-1\rangle.
\end{equation}

Now we consider a general initial state in the collective manifold, 
\begin{equation}
    |\psi_0\rangle=\sum_m c_m |N/2,m\rangle,
\end{equation}
with the constraints $\langle\psi_0|\hat{S}_n^{y}|\psi_0\rangle=0$, $\langle\psi_0|\hat{S}_n^{z}|\psi_0\rangle=0$. With time evolution under $\hat{H}_{\mathrm{eff}}$, we have
\begin{equation}
    \begin{aligned}
    |\psi(t)\rangle&=e^{-i\hat{H}_{\mathrm{eff}}t/\hbar}\\
    &\approx\sum_mc_m\bigg[e^{-iNJ_{\mathrm{eff}}t/2}|N/2,m\rangle-i\frac{\omega_{\mathrm{eff}}}{NJ_{\mathrm{eff}}}\sqrt{\frac{(N/2)^2-m^2}{N-1}}\sin(NJ_{\mathrm{eff}}t/2)|N/2-1,m\rangle\bigg],
    \end{aligned}
\end{equation}
which gives
\begin{equation}
    \begin{aligned}
    \langle \psi(t)|\hat{S}_j^{+}|\psi(t)\rangle&\approx \sum_m c_mc^{*}_{m+1}\bigg[\langle N/2,m+1|\hat{S}_j^{+}|N/2,m\rangle\\
    &-i\frac{\omega_{\mathrm{eff}}}{NJ_{\mathrm{eff}}}e^{iNJ_{\mathrm{eff}}t/2}\sqrt{\frac{(N/2)^2-m^2}{N-1}}\sin(NJ_{\mathrm{eff}}t/2)\langle N/2,m+1|\hat{S}_j^{+}|N/2-1,m\rangle\\
    &+i\frac{\omega_{\mathrm{eff}}}{NJ_{\mathrm{eff}}}e^{-iNJ_{\mathrm{eff}}t/2}\sqrt{\frac{(N/2)^2-(m+1)^2}{N-1}}\sin(NJ_{\mathrm{eff}}t/2)\langle N/2-1,m+1|\hat{S}_j^{+}|N/2,m\rangle\bigg]\\
    &=\sum_m c_mc^{*}_{m+1}\langle N/2,m+1|\hat{S}_j^{+}|N/2,m\rangle\bigg[1-(-1)^j\frac{\omega_{\mathrm{eff}}}{NJ_{\mathrm{eff}}}\bigg(i\frac{1}{2}\sin(NJ_{\mathrm{eff}}t)-\frac{2m+1}{N-1}\sin^2(NJ_{\mathrm{eff}}t/2)\bigg)\bigg]\\
    &=\langle\psi_0|\hat{S}_j^{+}|\psi_0\rangle\bigg[1-i(-1)^j\frac{\omega_{\mathrm{eff}}}{2}\frac{\sin(NJ_{\mathrm{eff}}t)}{NJ_{\mathrm{eff}}}\bigg]+\frac{\langle\psi_0|(\hat{S}_n^{+}\hat{S}^{z}+\hat{S}^{z}\hat{S}_n^{+})|\psi_0\rangle}{N-1}(-1)^j\frac{\omega_{\mathrm{eff}}}{NJ_{\mathrm{eff}}}\sin^2(NJ_{\mathrm{eff}}t/2).\\
    \end{aligned}
\end{equation}

Therefore, we can obtain the frequency for each of the two large spins,
\begin{equation}
    \omega_j(t)\approx (-1)^{j-1}\frac{\omega_{\mathrm{eff}}}{2}\bigg(\frac{\sin(NJ_{\mathrm{eff}}t)}{NJ_{\mathrm{eff}}t}-2\frac{\langle\psi_0|(\hat{S}^{y}\hat{S}^{z}+\hat{S}^{z}\hat{S}^{y})|\psi_0\rangle}{(N-1)\langle\psi_0|\hat{S}^{x}|\psi_0\rangle}\frac{\sin^2(NJ_{\mathrm{eff}}t/2)}{NJ_{\mathrm{eff}}t}\bigg),
\end{equation}
leading to the synchronization time
\begin{equation}
    NJ_{\mathrm{eff}}t_{\mathrm{syn}}=\pi-2\arctan\bigg[\frac{\langle\psi_0|(\hat{S}^{y}\hat{S}^{z}+\hat{S}^{z}\hat{S}^{y})|\psi_0\rangle}{(N-1)\langle\psi_0|\hat{S}^{x}|\psi_0\rangle}\bigg].
\end{equation}

In the following, we consider three different types of initial states:
\begin{itemize}
\item $|\psi_0\rangle=|+x\rangle^{\otimes N}$

In this case we have $\langle\psi_0|(\hat{S}^{y}\hat{S}^{z}+\hat{S}^{z}\hat{S}^{y})|\psi_0\rangle=0$, which gives
\begin{equation}
    NJ_{\mathrm{eff}}t_{\mathrm{syn},0}=\pi.
\end{equation}
Here we use $t_{\mathrm{syn},0}$ to label the synchronization time with this unentangled initial state. This result agrees with the prediction using the Holstein–Primakoff approximation in the previous subsection.

\item $|\psi_0\rangle=e^{-iQ\hat{S}^{z}\hat{S}^{z}/N}|+x\rangle^{\otimes N}$

In this case we have
\begin{equation}
    \langle\psi_0|(\hat{S}^{y}\hat{S}^{z}+\hat{S}^{z}\hat{S}^{y})|\psi_0\rangle=\frac{N(N-1)}{2}\sin(Q/N)\cos^{N-2}(Q/N),
\end{equation}
\begin{equation}
    \langle\psi_0|\hat{S}^{x}|\psi_0\rangle=\frac{N}{2}\cos^{N-1}(Q/N),
    \label{eq:squsx}
\end{equation}
which gives
\begin{equation}
    NJ_{\mathrm{eff}}t_{\mathrm{syn}}=\pi-2Q/N.
\end{equation}
Note that the $\hat{S}^z\hat{S}^z$ term commutes with $\hat{H}_{\mathrm{eff}}$, this result can also apply to the following Hamiltonian generated by cavity-mediated interactions,
\begin{equation}
    \hat{H}_{\mathrm{cav}}/\hbar=J_{\perp}\hat{\mathbf{S}}\cdot\hat{\mathbf{S}}+(J_z-J_{\perp})\hat{S}^z\hat{S}^z+\frac{\omega_{\mathrm{eff}}}{2}(\hat{S}^z_1-\hat{S}^z_2),
\end{equation}
with initial state $|\psi_0\rangle=|+x\rangle^{\otimes N}$. Now we can replace $J_{\rm eff}$ by $J_{\perp}$, and $Q/N$ by $(J_{z}-J_{\perp}) t_{\mathrm{syn}}$, which gives
\begin{equation}
    \Big((N-2)J_{\perp}+2J_z\Big)t_{\mathrm{syn}}=\pi.
\end{equation}
We have shown numerical calculations for collective spin exchange interactions ($J_z=0$) and collective Heisenberg interactions ($J_z=J_{\perp}$) with $N$ atoms in Fig.~\ref{fig:syntime}(c), which agrees well with the analytic results using two large spins.

\item $|\psi_0\rangle=e^{-i\theta\hat{S}^{x}}e^{-iQ\hat{S}^{z}\hat{S}^{z}/N}|+x\rangle^{\otimes N}$

In this case we have
\begin{equation}
    \langle\psi_0|(\hat{S}^{y}\hat{S}^{z}+\hat{S}^{z}\hat{S}^{y})|\psi_0\rangle=\cos(2\theta)\frac{N(N-1)}{2}\sin(Q/N)\cos^{N-2}(Q/N)+\sin(2\theta)\frac{N(N-1)}{8}\Big(1-\cos^{N-2}(2Q/N)\Big),
\end{equation}
and $\langle\psi_0|\hat{S}^{x}|\psi_0\rangle$ is still given by Eq.~(\ref{eq:squsx}). So the tunability of synchronization time is due to the $\theta$-rotation of $\langle\psi_0|(\hat{S}^{y}\hat{S}^{z}+\hat{S}^{z}\hat{S}^{y})|\psi_0\rangle$. In Fig.~4 of the main text, we find that the analytic results for two large spins can also explain the numerical simulations for collective or nearest-neighbor Heisenberg interactions, if we rewrite the formula of synchronization time $t_{\mathrm{syn}}$ and its tunable range $\Delta t_{\mathrm{syn}}$ into dimensionless form,
\begin{equation}
    \frac{t_{\mathrm{syn}}}{t_{\mathrm{syn},0}}=1-\frac{2}{\pi}\arctan\bigg[\frac{\langle\psi_0|(\hat{S}^{y}\hat{S}^{z}+\hat{S}^{z}\hat{S}^{y})|\psi_0\rangle}{(N-1)\langle\psi_0|\hat{S}^{x}|\psi_0\rangle}\bigg],
    \label{eq:ana1}
\end{equation}
\begin{equation}
    \frac{\Delta t_{\mathrm{syn}}}{t_{\mathrm{syn},0}}= \frac{4}{\pi}\arctan\bigg[\frac{(\langle\psi_0|(\hat{S}^{y}\hat{S}^{z}+\hat{S}^{z}\hat{S}^{y})|\psi_0\rangle_{\mathrm{max}}}{(N-1)\langle\psi_0|\hat{S}^{x}|\psi_0\rangle}\bigg].
    \label{eq:ana2}
\end{equation}
In the main text, we use $|\psi(\theta)\rangle$ instead of $|\psi_0\rangle$ in this case to emphasize the $\theta$-dependence.

\end{itemize}

\subsection{Experimental considerations}
Our discussions in the main text assume the frequency synchronization dynamics is purely unitary. Considering the existence of dissipative effects such as cavity photon loss, we would like to perform an order of magnitude analysis for the dissipative effects.

For simplicity, we analyze the case without the dressing scheme, so the $|\negmedspace\downarrow\rangle$ state is defined in the ${}^1S_0$ ground manifold of Sr, the $|\negmedspace\uparrow\rangle$ state is defined in the ${}^3P_0$ excited manifold, connected via an ultranarrow optical transition.
This setup is the same as Ref.~\cite{norcia2018cavity}, so the dynamics without dressing is captured by the following Lindblad master equation,
\begin{equation}
    \frac{d}{dt}\hat{\rho}=-i[J_{\perp}\hat{S}^{+}\hat{S}^{-}+\hat{H}_{\rm GRS},\hat{\rho}] + \Gamma\bigg(\hat{S}^{-}\hat{\rho}\hat{S}^{+}-\frac{1}{2}(\hat{S}^{+}\hat{S}^{-}\hat{\rho}+\hat{\rho}\hat{S}^{+}\hat{S}^{-})\bigg),
    \label{eq:exch}
\end{equation}
where $J_{\perp}=g_c^2\Delta_c/(\Delta_c^2+\kappa^2/4)$ is the spin exchange interaction strength, and $\Gamma=g_c^2\kappa/(\Delta_c^2+\kappa^2/4)$ is the superradiant decay rate (Lindblad jump operator $\hat{S}^{-}$ originated from cavity loss). 
Here $2g_c$ is the single-photon Rabi frequency, $\Delta_c$ is the cavity detuning with respect to the atomic transition, and $\kappa$ is the cavity linewidth.
The ratio $\Gamma/J_{\perp}$ is determined by
\begin{equation}
    \frac{\Gamma}{J_{\perp}} = \frac{\kappa}{\Delta_c}.
\end{equation}
In the regime where $\Delta_c \gg \kappa$, the dynamics can be considered as near-unitary dynamics in the time scale $N \Gamma  t \ll 1$.

If we use the same cavity as in Ref.~\cite{norcia2018cavity}, with $g_c/2\pi=4$Hz, and $\kappa/2\pi=160$kHz, we can set $\Delta_c/2\pi=5$MHz with $N=10^5$ atoms, leading to $NJ_{\perp}/2\pi=0.32$Hz, $N \Gamma/2\pi=0.01$Hz. Other dissipative effects such as spontaneous emission occurs at a rate $\gamma/2\pi \sim 1$mHz, which is smaller compared to superradiant decay. Including the dressing scheme, we will have an additional dissipative effect known as collective dephasing (Lindblad jump operator $\hat{S}^z$), which occurs at a comparable rate as the superradiant decay.

The frequency synchronization occurs at a time scale $NJ_{\perp}t_{\rm syn}/2\pi\sim 0.5 \;\Rightarrow\; t_{\rm syn} \sim 1.6$s, and one can check that $N \Gamma t_{\rm syn}\sim 0.1 \ll 1$. So it is possible to achieve frequency synchronization within $2$s, while in this time scale the dissipative effects remain subdominant. 
Dissipative effects would be more significant for highly entangled states.
At the synchronization time scale, a lower bound of spin squeezing can be estimated by $\min(\xi^2)\sim N \Gamma t_{\rm syn}\sim 0.1$. Therefore, the system dynamics is still approximately unitary for initial spin squeezed states below 10dB.
Note that all these estimations are based on experimental parameters in Ref.~\cite{norcia2018cavity}, and there is no fundamental limit in the experiment to reduce the cavity linewidth $\kappa$ and further suppress the cavity induced dissipative effects.